\documentclass[11pt]{article}

\usepackage[utf8]{inputenc}
\usepackage{graphicx}
\usepackage{longtable}
\usepackage{float}
\usepackage{listings}
\usepackage{tabularx}
\usepackage{xcolor}
\usepackage{pgfplots}
\pgfplotsset{compat=1.18}
\usepackage{tikz}
\usepackage{float}
\usepackage{amsmath}
\usepackage{amssymb}
\usepackage{booktabs}
\usepackage{hyperref}
\usepackage{cite}
\usepackage{subcaption}
\usepackage{algorithm}
\usepackage{algpseudocode}
\usepackage{pifont} % checkmarks / crosses

\definecolor{codegreen}{rgb}{0,0.6,0}
\definecolor{codegray}{rgb}{0.5,0.5,0.5}
\definecolor{codepurple}{rgb}{0.58,0,0.82}
\definecolor{backcolour}{rgb}{0.95,0.95,0.92}

\lstdefinestyle{mystyle}{
    backgroundcolor=\color{backcolour},
    commentstyle=\color{codegreen},
    keywordstyle=\color{magenta},
    numberstyle=\tiny\color{codegray},
    stringstyle=\color{codepurple},
    basicstyle=\ttfamily\footnotesize,
    breakatwhitespace=false,
    breaklines=true,
    captionpos=b,
    keepspaces=true,
    numbers=left,
    numbersep=5pt,
    showspaces=false,
    showstringspaces=false,
    showtabs=false,
    tabsize=2
}
\begin{document}

\title{Reliable Graph-RAG for Codebases: AST-Derived Graphs vs LLM-Extracted Knowledge Graphs}

\author{
Manideep Reddy Chinthareddy\\
Software Engineer, Centerville, USA \\
\texttt{chmanideepreddy@gmail.com}
}

\date{January 2026}
\maketitle

% ------------------------
% Abstract
% ------------------------
\begin{abstract}
Retrieval-Augmented Generation (RAG) for software engineering often relies on vector similarity search, which captures topical similarity but frequently fails on multi-hop architectural reasoning (e.g., controller $\rightarrow$ service $\rightarrow$ repository chains, interface-driven wiring, and inheritance-based behavior)~\cite{Lewis2020Retrieval,Gao2023RetrievalSurvey}.
This paper benchmarks three retrieval pipelines on Java codebases (detailed on Shopizer; additional runs on ThingsBoard and OpenMRS Core): (A) No-Graph Naive RAG (vector-only), (B) an LLM-Generated Knowledge Graph RAG (LLM-KB), and (C) a deterministic AST-derived Knowledge Graph RAG (DKB) built via Tree-sitter parsing with bidirectional traversal~\cite{TreeSitter2025}.

Using a fixed suite of 15 architecture and code-tracing queries per repository (The questions are same for all the 3 approaches while they differ per repository to make it relevant to the code bases), we report indexing overhead, query-time latency, corpus coverage signals, and end-to-end cost.
Across repositories, DKB builds its ontology graph in seconds (2.81s on Shopizer; 13.77s on ThingsBoard; 5.60s on OpenMRS Core), whereas LLM-KB requires substantially longer LLM-mediated graph generation (200.14s on Shopizer; 883.74s on ThingsBoard; 222.17s on OpenMRS Core).
Critically, LLM-KB exhibits probabilistic \emph{indexing incompleteness}: on Shopizer, the extraction log flags 377 files as \texttt{SKIPPED/MISSED by LLM}, yielding a per-file success rate of 0.688 (833/1210) and shrinking the embedded corpus to 3465 chunks (0.641 coverage vs.\ the No-Graph baseline of 5403 chunks), compared to deterministic indexing at 4873 chunks (0.902 coverage).
This incompleteness also reduces the extracted graph footprint: LLM-KB produces 842 nodes versus 1158 nodes for DKB (0.727 node coverage), even though all approaches scan the same discovered files; however, LLM-KB fails to produce structured records for a subset, and those files consequently do not contribute to embeddings/graph.
While batching strategies, prompt tuning, and retries can reduce omissions, completeness remains mediated by stochastic model behavior and schema-bound extraction failure modes in tool-mediated indexing pipelines~\cite{Singh2025AgenticSurvey,Hayes2024ProbExtract}.

We also quantify end-to-end execution cost (indexing + answering the full 15-question suite) and report relative cost normalized to the No-Graph baseline.
On Shopizer, representative costs were \$0.04 (No-Graph), \$0.09 (DKB), and \$0.79 (LLM-KB), corresponding to $\sim$2.25$\times$ for DKB and $\sim$19.75$\times$ for LLM-KB.
On a larger combined OpenMRS-core + ThingsBoard workload, costs increase to \$0.149 (No-Graph), \$0.317 (DKB), and \$6.80 (LLM-KB), yielding $\sim$2.13$\times$ for DKB and $\sim$45.64$\times$ for LLM-KB, indicating that LLM-mediated graph construction can dominate total cost as repository scale increases.

For query-time latency, the No-Graph baseline remains competitive but less stable across workloads: on Shopizer it achieves 9.52$\pm$2.98s (mean$\pm$std), compared to 10.51$\pm$4.17s for DKB and 13.36$\pm$7.87s for LLM-KB.
On ThingsBoard, mean latencies increase to 10.92$\pm$1.43s (No-Graph), 11.17$\pm$1.97s (DKB), and 15.29$\pm$4.94s (LLM-KB), while on OpenMRS Core we observe 11.82$\pm$2.79s (No-Graph), 12.21$\pm$6.96s (DKB), and 11.94$\pm$4.88s (LLM-KB), with LLM-KB and DKB showing higher worst-case outliers in several settings.
We additionally validate answer correctness on the full 15-question Shopizer suite: DKB attains the highest correctness (15/15), LLM-KB follows closely (13/15; 2 partial), while No-Graph degrades on upstream architectural queries and exhibits the highest hallucination risk (6/15). Across repositories, correctness gains are largest on suites that emphasize multi-hop architectural tracing and upstream discovery; in some suites (e.g., ThingsBoard) DKB ties the vector-only baseline while maintaining higher coverage and lower indexing overhead than LLM-mediated graph construction.

\end{abstract}

\noindent\textbf{Keywords:}
Retrieval-Augmented Generation (RAG), Static Code Analysis, Abstract Syntax Trees (AST), Knowledge Graphs, Multi-Hop Reasoning, Software Maintenance.

% ------------------------
% Nomenclature
% ------------------------
\section*{Nomenclature}
\noindent
\begin{tabularx}{\columnwidth}{@{}l X@{}}
$G=(V,E)$ & Code knowledge graph with nodes $V$ and edges $E$ \\
$q$ & Natural language query \\
$R(\cdot)$ & Vector retriever returning top-$k$ items \\
$N_d(v)$ & Graph neighborhood within depth $d$ around node $v$ \\
$\mathcal{C}(q)$ & Context assembled for query $q$ \\
$Q$ & Number of benchmark questions (here $Q=15$) \\
$C$ & Number of embedded text chunks \\
$N_{\text{files}}$ & Number of Java files discovered during scanning \\
\end{tabularx}

% ------------------------
% Introduction
% ------------------------
\section{Introduction}
As Large Language Models (LLMs) transition from generic assistants to enterprise reasoning engines built on Transformer architectures~\cite{Vaswani2017Attention}, the context window remains a primary constraint.
Retrieval-Augmented Generation (RAG) mitigates this constraint by retrieving relevant context from a codebase prior to answer generation~\cite{Lewis2020Retrieval,Gao2023RetrievalSurvey}.
In software engineering settings, however, vector similarity often introduces \emph{context flattening}: the retrieved chunks share lexical or semantic overlap with the query, but do not reliably preserve structural dependencies such as inheritance, dependency injection, and call relationships.

In complex systems, many questions are inherently multi-hop.
For example, answering ``Which controllers use the shopping cart logic?{}`` requires traversing upstream consumers from services to controllers, and frequently crossing interface boundaries.
Vector-only retrieval may retrieve the shopping-cart implementation class but omit the controllers that depend on it, producing incomplete or ungrounded answers.
This failure mode is consistent with known limitations of dense retrieval pipelines (e.g., DPR-style retrievers) when evidence is distributed across multiple non-local contexts~\cite{Karpukhin2020Dense,Izacard2022Atlas}.

This paper compares three retrieval paradigms for code analysis:
(A) No-Graph Naive RAG (vector-only),
(B) LLM-KB Graph RAG (LLM-generated dependency graph during indexing),
and (C) DKB, a deterministic compiler-inspired approach that parses code via ASTs and performs bidirectional graph expansion at query time~\cite{TreeSitter2025,Zhang2025CAST}.
Our motivation aligns with repository-level code understanding needs highlighted by recent benchmark suites and surveys~\cite{Jimenez2023SWEbench,Tian2023RepoBench,Tao2025RetrievalSurvey,Yang2025Empirical}.

\subsection{Contributions}
This paper makes the following contributions:
\begin{enumerate}
  \item \textbf{Benchmarking framework for graph-aware retrieval:} We provide an end-to-end comparison of vector-only, LLM-extracted graph RAG, and AST-derived graph RAG under shared hyperparameters and identical question sets~\cite{Gao2023RetrievalSurvey,OpenReview2025GraphSurvey}.
  \item \textbf{Measured indexing and query-time costs:} We report concrete build times and latency distributions; notably, LLM-based graph construction introduces large indexing overhead compared to deterministic AST parsing~\cite{Edge2024GraphRAG,TreeSitter2025}.
  \item \textbf{Indexing reliability analysis:} We instrument and report \emph{coverage/consistency} signals from run logs (files scanned, nodes/edges built, chunks embedded), and show that LLM-KB can skip files during extraction, shrinking both graph size and embedded corpus; this connects to broader reliability discussions in agentic RAG and tool-mediated pipelines~\cite{Singh2025AgenticSurvey}.
  \item \textbf{Bidirectional + interface-aware expansion:} We show how predecessor traversal and interface-consumer expansion improve retrieval for architecture discovery queries that often fail under successors-only or text-only retrieval~\cite{Edge2024GraphRAG,OpenReview2025GraphSurvey}.
  \item \textbf{Correctness validation:} We provide a correctness comparison on representative questions spanning repository-level reasoning and multi-hop architectural discovery~\cite{Jimenez2023SWEbench,Tian2023RepoBench,Zhang2023RepoEval}.
\end{enumerate}

\subsection{Paper Organization}
Section II reviews background on RAG and program graphs.
Section III formalizes the problem and research questions.
Section IV details the three pipelines and the graph-aware retrieval algorithm.
Section V describes the experimental setup and metrics.
Section VI reports quantitative and qualitative results.
Sections VII--X discuss implications, threats to validity, reproducibility, and limitations before concluding.

% ------------------------
% Background / Preliminaries
% ------------------------
\section{Background and Preliminaries}

\subsection{RAG for Software Engineering}
In code intelligence, a common baseline is chunk-based embedding retrieval: source code is split into overlapping segments, embedded, and retrieved by similarity to a query~\cite{Lewis2020Retrieval,Gao2023RetrievalSurvey}.
This approach is simple and fast, but can fail to include structurally necessary context when the required evidence resides in related files (e.g., an interface definition, a parent class, or a controller).
Recent work on lightweight retrieval stacks emphasizes simplicity and speed, but does not eliminate the need for structured multi-hop grounding in codebases~\cite{Guo2024LightRAG}.

\subsection{Program Structure Graphs}
Code structure can be represented as graphs: nodes correspond to entities such as classes, interfaces, methods, and files; edges represent relations such as \textit{extends}, \textit{implements}, \textit{injects/uses}, and \textit{calls}.
Static parsing (e.g., AST analysis) can deterministically recover many such relations for statically-typed languages like Java~\cite{TreeSitter2025}.
AST-derived structure has also been used to improve code chunking for retrieval by aligning chunk boundaries with syntactic structure rather than raw character windows~\cite{Zhang2025CAST}.

\subsection{Knowledge Graphs and Ontologies in Code}
In this work, ``ontology'' refers to a typed schema for code entities and relations (node types + edge labels) suitable for retrieval and traversal.
Graph RAG leverages this structure to expand retrieved context beyond the initial top-$k$ chunks~\cite{Edge2024GraphRAG,Knollmeyer2025DocumentGraphRAG,OpenReview2025GraphSurvey}.

% ------------------------
% Related Work (fill with citations later)
% ------------------------
\section{Related Work}
% TODO: Add citations and positioning:
% (i) RAG for LLMs and code assistants,
% (ii) GraphRAG / KG-RAG systems,
% (iii) LLM-based code graph extraction,
% (iv) static analysis and program graphs,
% (v) hybrid compiler+LLM systems.

\textbf{Retrieval-augmented generation.}
RAG was formalized as a paradigm for grounding sequence generation in retrieved evidence, combining parametric and non-parametric memory to improve factuality and reduce hallucinations~\cite{Lewis2020Retrieval}.
Dense retrievers such as DPR established effective neural retrieval for open-domain QA, and subsequent retrieval-augmented language models (e.g., ATLAS) showed strong few-shot and knowledge-intensive performance by tightly coupling retrieval with generation~\cite{Karpukhin2020Dense,Izacard2022Atlas}.
Recent surveys consolidate best practices in retrieval pipelines, chunking, indexing, and evaluation, while noting that retrieval adequacy remains workload-dependent and can degrade under multi-hop evidence requirements~\cite{Gao2023RetrievalSurvey}.

\textbf{GraphRAG and knowledge-graph enhanced RAG.}
GraphRAG-style systems introduce explicit graph structure to move beyond purely local similarity matches.
Edge et al.\ propose a ``local-to-global'' GraphRAG approach for query-focused summarization, where initial retrieval is expanded through graph neighborhoods to collect globally relevant supporting context~\cite{Edge2024GraphRAG}.
Domain-focused variants, such as Document GraphRAG in manufacturing QA, demonstrate that knowledge-graph signals can improve document QA by connecting entities and relations across otherwise disjoint textual fragments~\cite{Knollmeyer2025DocumentGraphRAG}.
A recent survey of graph retrieval-augmented generation discusses a design space spanning graph construction (manual vs.\ extracted), graph granularity (document vs.\ entity), and graph-guided retrieval policies~\cite{OpenReview2025GraphSurvey}.
Our work differs by targeting \emph{code} graphs where structural relations are not merely semantic co-occurrence but compiler-visible dependencies.

\textbf{RAG for code and repository-level evaluation.}
LLMs trained on code enabled strong program synthesis and code completion behavior, and evaluation frameworks for code LLMs highlight the importance of realistic, repository-grounded tasks~\cite{Chen2021Evaluating,Austin2021Program}.
Repository-level benchmarks such as SWE-bench and RepoBench emphasize multi-file reasoning, tool use, and integration with real code artifacts rather than single-function problems~\cite{Jimenez2023SWEbench,Tian2023RepoBench}.
RepoEval (introduced in repository-level completion work) similarly focuses on evaluating completion and generation in the presence of repository context~\cite{Zhang2023RepoEval}.
Surveys specifically focused on retrieval-augmented code generation summarize repository-level approaches and emphasize that retrieval quality, chunking strategy, and grounding across file boundaries are often the limiting factors~\cite{Tao2025RetrievalSurvey,Yang2025Empirical}.

\textbf{Structure-aware chunking and code graphs.}
Structure-aware chunking using ASTs (e.g., cAST/CAST) improves retrieval by preventing semantically coupled code from being split across chunks and by avoiding malformed fragments that degrade embedding utility~\cite{Zhang2025CAST}.
Graph-based retrieval methods for repository-level code generation, such as RepoGraph, explicitly leverage repository structure to retrieve and compose evidence across files, aligning with the core motivation of this paper~\cite{He2024RepoGraph}.
Beyond retrieval, graph-integrated models (e.g., CGM) represent a complementary line of work that integrates graph signals into model architectures for repository-level software engineering tasks~\cite{Liu2025CodeGraphModel}.

\textbf{Determinism, robustness, and indexing reliability.}
While many GraphRAG systems assume a stable graph artifact, LLM-mediated graph extraction can introduce stochasticity and schema-compliance failures, and these effects are amplified in agentic RAG settings where multiple tool calls and intermediate steps can compound variance~\cite{Singh2025AgenticSurvey}.
In practice, this may manifest as missing entities/edges or skipped inputs during indexing, producing blind spots at query time; our contribution is to measure this effect directly via run-log coverage signals in a codebase setting.
More broadly, when the downstream task is correctness-sensitive (e.g., vulnerability detection), systematic benchmarks have found meaningful gaps between LLM-driven approaches and traditional static analysis under controlled evaluation, reinforcing the value of deterministic signals where available~\cite{LLMvsStatic2025}.

% ------------------------
% Problem Formulation + RQs
% ------------------------
\section{Problem Formulation and Research Questions}

\subsection{Task Definition}
Let $\mathcal{S}$ be a codebase, and let $q$ be a natural-language query about code behavior or architecture.
A RAG system assembles context $\mathcal{C}(q)$ and produces an answer $a$.
We study retrieval strategies that differ in whether and how they represent code structure~\cite{Lewis2020Retrieval,Gao2023RetrievalSurvey}.

\subsection{Graph-Based Retrieval Formalization}
We model code structure as a directed labeled graph $G=(V,E)$, where nodes represent code entities (classes, interfaces, etc.) and edges represent relations (e.g., \textit{extends}, \textit{implements}, \textit{injects})~\cite{TreeSitter2025,OpenReview2025GraphSurvey}.

A graph-aware retriever expands an initially retrieved node set $V_0$ using neighborhood expansion:
\begin{equation}
V_d = V_0 \cup \bigcup_{v \in V_0} N_d(v),
\end{equation}
where $N_d(v)$ denotes the set of nodes within hop depth $d$ around node $v$ under selected edge directions (successors, predecessors, or both)~\cite{Edge2024GraphRAG}.

The final assembled context can be expressed as:
\begin{equation}
\mathcal{C}(q) = \text{Concat}\left(\{\text{code}(v) \mid v \in V_d\}\right).
\end{equation}

\subsection{Research Questions}
\begin{itemize}
  \item \textbf{RQ1 (Efficiency):} What are the indexing time and query latency trade-offs across vector-only, LLM-KB, and DKB?~\cite{Gao2023RetrievalSurvey}
  \item \textbf{RQ2 (Retrieval adequacy):} Does deterministic bidirectional expansion retrieve evidence that vector-only retrieval misses on multi-hop questions?~\cite{Edge2024GraphRAG,OpenReview2025GraphSurvey}
  \item \textbf{RQ3 (Stability):} How variable is each approach in query-time latency, and does LLM-KB introduce higher variance or worst-case outliers due to multi-step/tool-mediated processing?~\cite{Singh2025AgenticSurvey}
  \item \textbf{RQ4 (Correctness):} Which approach produces the most accurate answers across representative code comprehension and architecture discovery tasks?~\cite{Jimenez2023SWEbench,Tian2023RepoBench}
  \item \textbf{RQ5 (Indexing reliability):} How often does each approach preserve corpus coverage (files, entities, chunks), and does LLM-KB exhibit extraction-time skipping that reduces downstream retrievability?
\end{itemize}

% ------------------------
% System Overview + Graph Artifacts
% ------------------------
\section{System Overview}

\subsection{Compared Pipelines}
We compare three pipelines:
\begin{enumerate}
  \item \textbf{Naive RAG (No-Graph):} top-$k$ vector similarity retrieval over code chunks~\cite{Lewis2020Retrieval,Karpukhin2020Dense}.
  \item \textbf{LLM-KB:} LLM-generated dependency graph at indexing time, plus graph-informed context expansion at query time~\cite{Edge2024GraphRAG,OpenReview2025GraphSurvey}.
  \item \textbf{DKB (Ours):} deterministic AST-derived graph built with Tree-sitter, used for bidirectional traversal and interface-aware expansion at query time~\cite{TreeSitter2025,Zhang2025CAST}.
\end{enumerate}

\subsection{Graph Artifacts}
\begin{figure*}[t]
\centering
\begin{subfigure}{0.49\textwidth}
\includegraphics[width=\linewidth]{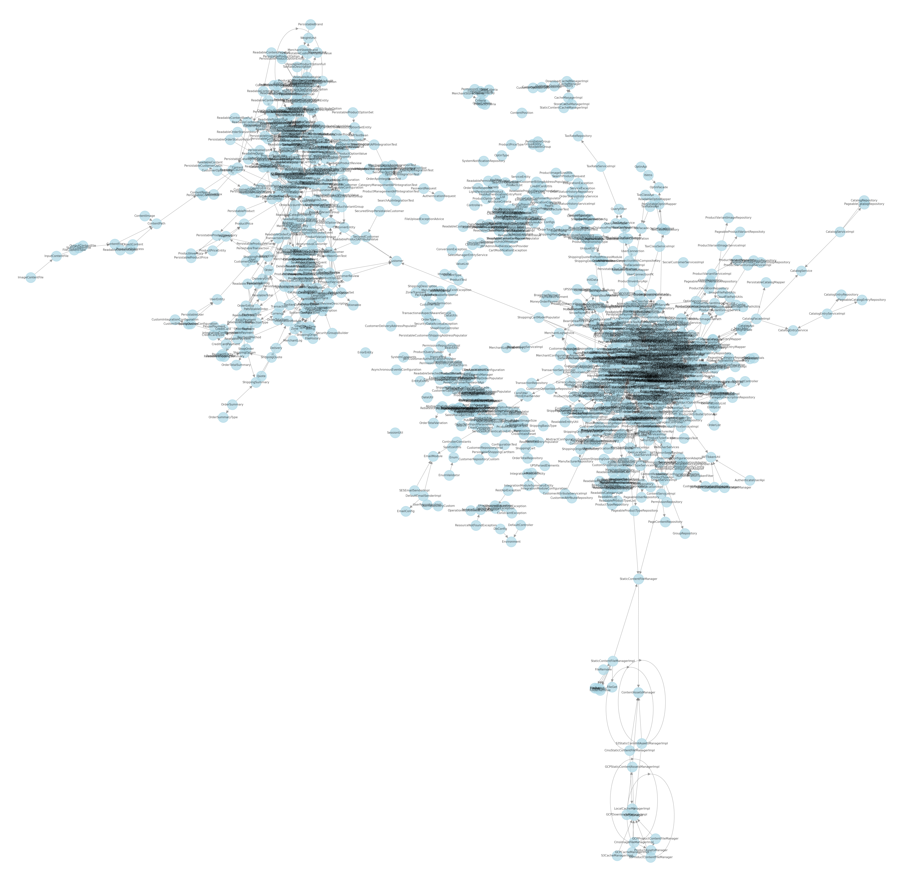}
\caption{Deterministic AST-derived ontology graph (DKB).}
\end{subfigure}
\hfill
\begin{subfigure}{0.49\textwidth}
\includegraphics[width=\linewidth]{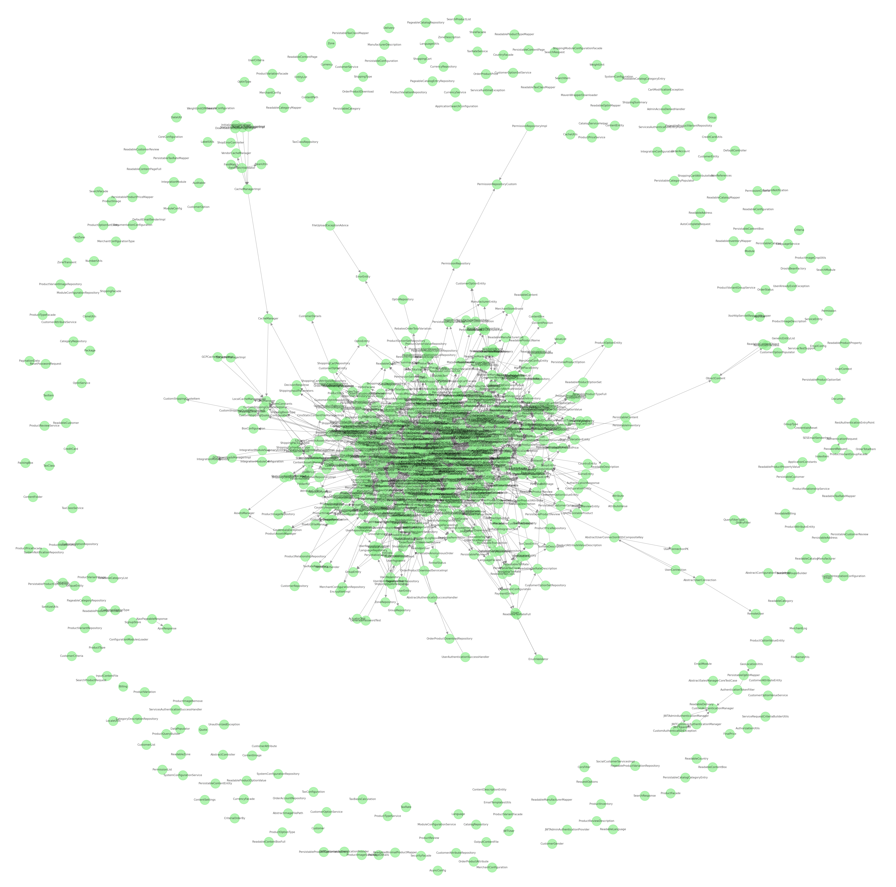}
\caption{LLM-extracted dependency graph (LLM-KB).}
\end{subfigure}
\caption{Graph artifacts generated for the same Java codebase (Shopizer). \textbf{DKB:} nodes are project-local Java types (classes/interfaces/enums/records/annotations) discovered via Tree-sitter; edges are typed \texttt{injects}, \texttt{extends}, and \texttt{implements} relations. \textbf{LLM-KB:} nodes are LLM-emitted \texttt{class\_name} entities; edges are \texttt{depends\_on} relations extracted from per-file structured outputs after filtering out standard-library and framework dependencies (e.g., \texttt{java.*}, Spring packages). (This caption replaces the earlier placeholder request to specify node/edge definitions and filtering.)}
\label{fig:graphs}
\end{figure*}

Figure~\ref{fig:graphs} shows the two graph artifacts produced for the same codebase.

% ------------------------
% Methodology
% ------------------------
\section{Methodology}

\subsection{Strategy A: Naive Vector Retrieval (No-Graph)}
We implement a standard RAG baseline that relies exclusively on vector similarity search~\cite{Lewis2020Retrieval,Karpukhin2020Dense}.
Source files are chunked using a recursive character splitter (\texttt{chunk\_size=1000}, \texttt{chunk\_overlap=100}), embedded, and indexed in a local vector store.
At query time, the system retrieves the top-$k=10$ chunks by similarity and passes them to an LLM for answer generation.

\subsection{Strategy B: LLM-Generated Knowledge Graph (LLM-KB)}
The LLM-KB approach constructs a dependency graph by prompting an LLM during indexing, aligned with graph-based retrieval-augmentation patterns in the broader GraphRAG literature~\cite{Edge2024GraphRAG,OpenReview2025GraphSurvey}.
For each Java file, the LLM emits a structured JSON object containing entities and dependencies.
A directed graph is then created and used to expand context at query time by appending neighboring nodes and their code content.

\textbf{Practical constraint.} To fit batch prompts in context limits, the implementation truncates file content when building per-batch inputs. This can contribute to partial extraction or omission of a file from the structured output, which the run logs surface as ``SKIPPED/MISSED by LLM''.
While stricter prompting, smaller batch sizes, or retries can reduce the skip rate, complete coverage is not guaranteed in multi-step/agentic retrieval pipelines where intermediate tool outputs must satisfy schemas and control-flow checks~\cite{Singh2025AgenticSurvey}.

\subsection{Strategy C: Deterministic AST Graph (DKB - Proposed)}
DKB deterministically constructs a code graph using Tree-sitter (\texttt{tree\_sitter\_java})~\cite{TreeSitter2025}.
The pipeline adds labeled edges for inheritance (\texttt{extends}/\texttt{implements}) and dependency injection patterns (field and constructor parameter types).
At query time, DKB expands context using \emph{bidirectional} graph traversal (successors and predecessors), and performs \textbf{interface-consumer expansion} to improve upstream discovery.
This design is complementary to structure-aware chunking approaches (e.g., AST-based chunking) that improve retrieval signal quality by aligning chunks with syntactic units~\cite{Zhang2025CAST}.

\subsection{Graph-Aware Retrieval Algorithm}
\begin{algorithm}[H]
\caption{Graph-aware context assembly (bidirectional)}
\label{alg:graph_rag}
\begin{algorithmic}[1]
\Require query $q$, vector retriever $R$, graph $G$, depth $d$, top-$k$
\State $D \gets R(q)$ \Comment{Top-$k$ retrieved chunks}
\State $V_0 \gets \textsc{EntitiesFromDocs}(D)$
\State $V \gets V_0$
\For{$v \in V_0$}
  \State $V \gets V \cup \textsc{Succ}(G,v,d) \cup \textsc{Pred}(G,v,d)$
  \State $V \gets V \cup \textsc{InterfaceConsumerExpand}(G,v)$ \Comment{Optional}
\EndFor
\State \Return $\mathcal{C}(q) \gets \textsc{AssembleCodeContext}(V)$
\end{algorithmic}
\end{algorithm}

% ------------------------
% Implementation Snippets (ADDED; existing content unchanged)
% ------------------------
\subsection{Implementation Snippets and Instrumentation}
This section provides small excerpts from the evaluation scripts to make the indexing and retrieval behaviors concrete and reproducible.

\subsubsection{DKB: Deterministic ontology extraction (Tree-sitter)}
Listing~\ref{lst:dkb_ast_extract} shows the Tree-sitter query patterns and the two-pass extraction strategy used to build an AST-derived graph with typed edges (\texttt{injects}, \texttt{extends}, \texttt{implements})~\cite{TreeSitter2025}.

\begin{lstlisting}[language=Python, caption={DKB ontology construction via Tree-sitter queries and two-pass extraction.}, label={lst:dkb_ast_extract}]
# Tree-sitter queries for type discovery + DI signals
class_query = Query(JAVA_LANGUAGE, """
(class_declaration name: (identifier) @class_name)
(interface_declaration name: (identifier) @class_name)
(enum_declaration name: (identifier) @class_name)
(record_declaration name: (identifier) @class_name)
(annotation_type_declaration name: (identifier) @class_name)
""")

injection_query = Query(JAVA_LANGUAGE, """
(field_declaration type: (type_identifier) @type_name)
""")

constructor_query = Query(JAVA_LANGUAGE, """
(constructor_declaration parameters: (formal_parameters
  (formal_parameter type: (type_identifier) @type_name)))
""")

def build_spring_graph_treesitter(root_path: str):
    G = nx.DiGraph()
    file_map = {}

    # Pass 1: discover project types -> file map
    for root, _, files in os.walk(root_path):
        for file in files:
            if not file.endswith(".java"): continue
            full_path = os.path.join(root, file)
            content = open(full_path, "rb").read()
            tree = parser.parse(content)
            captures = class_cursor.captures(tree.root_node)
            for capture_name, nodes in captures.items():
                for node in nodes:
                    if capture_name == "class_name":
                        class_name = get_node_text(node, content)
                        file_map[class_name] = full_path
                        G.add_node(class_name, path=full_path)

    # Pass 2: add edges: injects / extends / implements
    for class_name, file_path in file_map.items():
        content = open(file_path, "rb").read()
        tree = parser.parse(content)

        # Field + constructor injections
        for _, nodes in injection_cursor.captures(tree.root_node).items():
            for node in nodes:
                dep = get_node_text(node, content)
                if dep in file_map: G.add_edge(class_name, dep, relation="injects")

        for _, nodes in constructor_cursor.captures(tree.root_node).items():
            for node in nodes:
                dep = get_node_text(node, content)
                if dep in file_map: G.add_edge(class_name, dep, relation="injects")

        # Inheritance edges (extends/implements) extracted by field names
        # ... (omitted for brevity; see full script)
    return G, file_map
\end{lstlisting}

\subsubsection{DKB: Bidirectional traversal + interface-consumer expansion}
Listing~\ref{lst:dkb_bidirectional} shows the \emph{bidirectional} graph expansion used at query time, including the key interface-consumer fix that improves controller discovery across interface boundaries~\cite{OpenReview2025GraphSurvey}.

\begin{lstlisting}[language=Python, caption={DKB query-time context assembly with successors, predecessors, and interface-consumer expansion.}, label={lst:dkb_bidirectional}]
def retrieve_with_graph_context(query: str) -> str:
    docs = retriever.invoke(query)
    context_text = ""
    processed_classes = set()

    for doc in docs:
        source_path = doc.metadata.get("source", "")
        class_name = os.path.basename(source_path).replace(".java", "")

        if class_name in processed_classes: continue
        processed_classes.add(class_name)

        # Bidirectional expansion
        if class_name in graph:
            successors = list(graph.successors(class_name))      # downstream deps
            predecessors = list(graph.predecessors(class_name))  # upstream users

            # INTERFACE EXPANSION:
            # if class implements an interface, also add consumers of that interface
            for successor in successors:
                edge_data = graph.get_edge_data(class_name, successor)
                if edge_data and edge_data.get("relation") == "implements":
                    interface_users = list(graph.predecessors(successor))
                    predecessors.extend(interface_users)
                    context_text += (
                      f"\n[ONTOLOGY INFO]: {class_name} implements {successor}. "
                      f"Checking consumers of {successor}...\n"
                    )

            # Emit relationship summary and append code of neighbors (budgeted)
            context_text += f"\n[ONTOLOGY INFO]: Relationships for {class_name}:\n"
            for dep in successors:
                rel = graph.get_edge_data(class_name, dep).get("relation", "uses")
                context_text += f"  - [INJECTS/USES] -> {dep} ({rel})\n"

            for consumer in predecessors:
                rel = (graph.get_edge_data(consumer, class_name) or {}).get(
                    "relation", "uses (via interface)"
                )
                context_text += f"  - [USED BY] <- {consumer} ({rel})\n"

    return context_text
\end{lstlisting}

\subsubsection{LLM-KB: Batch extraction, truncation, and explicit skip detection}
Listing~\ref{lst:llmkb_skip} highlights the probabilistic indexing behavior: file contents are truncated for batching, and the pipeline prints explicit \texttt{SKIPPED/MISSED by LLM} lines when a file in the input batch does not appear in the model’s structured output; this kind of schema-bound omission is a common reliability concern in agentic/tool-mediated RAG pipelines, and more broadly reflects known limitations and trade-offs in enforcing strict structured outputs in practice~\cite{Singh2025AgenticSurvey,Geng2025StructuredOutputs,Agarwal2025ThinkJSON}.

\begin{lstlisting}[language=Python, caption={LLM-KB batch analysis with file truncation and explicit skipped-file detection.}, label={lst:llmkb_skip}]
# Batch preparation: truncate each file to fit prompt budget
for path in batch_files:
    content = open(path, "r", encoding="utf-8").read()
    snippet = content[:15000]  # truncation to save context
    batch_input_str += f"\n--- FILE: {path} ---\n{snippet}\n"

# Invoke LLM for the whole batch (structured JSON)
result = extraction_chain.invoke({"batch_content": batch_input_str})
analyzed_files = result.get("results", [])

# Track which input paths appeared in structured output
processed_paths_in_batch = set()
for res in analyzed_files:
    c_path = (res.get("file_path") if isinstance(res, dict) else res.file_path)
    if c_path:
        processed_paths_in_batch.add(os.path.normpath(c_path))

# Emit explicit skip signals if an input file is missing from output
for original_path in input_batch:
    norm_path = os.path.normpath(original_path)
    if norm_path not in processed_paths_in_batch:
        # fallback: basename match (LLM may return relative paths)
        base_name = os.path.basename(norm_path)
        found = any(os.path.basename(p) == base_name for p in processed_paths_in_batch)
        if not found:
            print(f"SKIPPED/MISSED by LLM: {os.path.basename(norm_path)}")
\end{lstlisting}

% ------------------------
% Experimental Setup
% ------------------------
\section{Experimental Setup}
\subsection{Dataset}
We evaluate generalization across three Java codebases with distinct architectures and domains:
(i) \textbf{Shopizer} (Spring-based e-commerce),
(ii) \textbf{ThingsBoard} (IoT platform with transport + actor/rule-engine subsystems),
and (iii) \textbf{OpenMRS Core} (clinical EMR platform with web filters, service layer, and modular runtime).
For each repository, we run the same 15-question suite under No-Graph RAG, LLM-KB (LLM-built graph), and DKB (Tree-sitter graph).

\subsection{Task Set}
We use a fixed suite of $Q=15$ architecture and code-tracing questions for each of the 3 approaches. Note that the question suite is different for different repositories. This is done to ensure the questions stay relevant to the repository.
This paper reports detailed correctness for the full 15-question suite spanning repository semantics, multi-hop service tracing, event-driven triggers, upstream discovery, and ID generation.
The emphasis on multi-file, repository-level reasoning mirrors recent evaluation directions in software engineering benchmarks~\cite{Jimenez2023SWEbench,Tian2023RepoBench,Zhang2023RepoEval}.

% ------------------------
% Evaluation Metrics
% ------------------------
\section{Evaluation Metrics}

\subsection{Latency}
We report mean, standard deviation, median, and min--max of per-question latency:
\begin{equation}
\overline{T} = \frac{1}{Q}\sum_{i=1}^{Q} T_i.
\end{equation}

\subsection{Correctness Labels}
We label each answer as \textbf{Correct}, \textbf{Partial}, or \textbf{Incorrect} based on whether it matches the repository’s ground-truth behavior/structure and avoids hallucinated entities.
For tabular summary, we collapse \emph{Correct+} detail level into \emph{Correct}.
This evaluation framing is consistent with repository-level code tasks that require grounded behavior and multi-file evidence~\cite{Jimenez2023SWEbench,Tao2025RetrievalSurvey,Yang2025Empirical}.

\subsection{Indexing Reliability and Coverage}
In addition to time, we measure \emph{indexing reliability} using run-log signals indicating how much of the corpus was successfully indexed.

\textbf{Chunk coverage.} Let $C_{\text{baseline}}$ be the number of embedded chunks produced by the vector-only pipeline over all discovered Java files, and let $C_{\text{approach}}$ be the number of chunks embedded under a given approach. Then:
\begin{equation}
\text{ChunkCoverage} = \frac{C_{\text{approach}}}{C_{\text{baseline}}}.
\end{equation}

\textbf{Graph coverage.} For graph-based methods, we report graph size ($|V|$, $|E|$) as a proxy for entity coverage. When comparing DKB vs.\ LLM-KB on the same codebase, we also report:
\begin{equation}
\text{NodeCoverage (LLM-KB vs.\ DKB)} = \frac{|V|_{\text{LLM-KB}}}{|V|_{\text{DKB}}}.
\end{equation}

\textbf{Skip indicators.} LLM-KB emits explicit log lines of the form ``\texttt{SKIPPED/MISSED by LLM: \textless File.java\textgreater}'' whenever a file from the input batch does not appear in the model's structured output. We treat this as direct evidence of probabilistic extraction incompleteness in a multi-step, tool-mediated indexing pipeline; this is consistent with broader findings that strict structured-output requirements can fail or degrade under realistic schema constraints and generation setups~\cite{Singh2025AgenticSurvey,Geng2025StructuredOutputs,Agarwal2025ThinkJSON}.

\textbf{Per-file success rate.} To emphasize extraction stochasticity, we also report the fraction of discovered files that were successfully processed into structured records (and therefore eligible for embedding/graph insertion). Let $N_{\text{processed}}$ denote the number of files that produced an analyzed record. We define:
\begin{equation}
\text{FileSuccessRate} = \frac{N_{\text{processed}}}{N_{\text{files}}}.
\end{equation}
For LLM-KB, we compute $N_{\text{processed}} = N_{\text{files}} - N_{\text{skipped}}$, where $N_{\text{skipped}}$ is the number of \emph{unique} filenames flagged by the log indicator \texttt{SKIPPED/MISSED by LLM}.

% ------------------------
% Results
% ------------------------
\section{Results}

\subsection{Quantitative Performance Benchmarks}
\begin{table*}[t]
\caption{Indexing time breakdown.}
\label{tab:indexing_time_json}
\centering
\small
\setlength{\tabcolsep}{6pt}
\renewcommand{\arraystretch}{1.10}
\begin{tabular}{llrrr}
\toprule
\textbf{Repository} & \textbf{Approach} & \textbf{DB time (s)} & \textbf{Graph time (s)} & \textbf{Total (s)} \\
\midrule
Shopizer     & No-Graph   & 18.41  & 0.00   & 18.41 \\
Shopizer     & DKB (Ours) & 19.28  & 2.81   & 22.09 \\
Shopizer     & LLM-KB     & 14.95  & 200.14 & 215.09 \\
\midrule
ThingsBoard  & No-Graph   & 123.61 & 0.00   & 123.61 \\
ThingsBoard  & DKB (Ours) & 129.78 & 13.77  & 143.55 \\
ThingsBoard  & LLM-KB     & 95.51  & 883.74 & 979.25 \\
\midrule
OpenMRS Core & No-Graph   & 42.20  & 0.00   & 42.20 \\
OpenMRS Core & DKB (Ours) & 42.82  & 5.60   & 48.42 \\
OpenMRS Core & LLM-KB     & 29.12  & 222.17 & 251.29 \\
\bottomrule
\end{tabular}
\end{table*}

\paragraph{Interpreting DB build time under corpus shrinkage.}
In Table~\ref{tab:indexing_time_json}, LLM-KB shows lower vector DB build time than No-Graph on some repositories.
This should not be interpreted as a more efficient embedding pipeline: because LLM-KB can omit files during structured extraction, it often embeds a smaller corpus (fewer chunks) than the No-Graph baseline (Tables~\ref{tab:coverage}, \ref{tab:thingsboard_coverage}, \ref{tab:openmrs_coverage}), which directly reduces DB construction work.
Accordingly, we treat \emph{total indexing time} (DB + graph) together with coverage signals as the relevant efficiency comparison.

\begin{table*}[t]
\caption{Query-time latency over the $Q=15$ question suite (from JSON run artifacts).}
\label{tab:query_latency_json}
\centering
\small
\setlength{\tabcolsep}{6pt}
\renewcommand{\arraystretch}{1.10}
\begin{tabular}{llrrr}
\toprule
\textbf{Repository} & \textbf{Approach} & \textbf{Mean$\pm$Std (s)} & \textbf{Median (s)} & \textbf{Min--Max (s)} \\
\midrule
Shopizer     & No-Graph   & 9.52$\pm$2.98  & 9.68  & 2.85--14.16 \\
Shopizer     & DKB (Ours) & 10.51$\pm$4.17 & 9.95  & 3.47--21.16 \\
Shopizer     & LLM-KB     & 13.36$\pm$7.87 & 10.96 & 6.53--38.25 \\
\midrule
ThingsBoard  & No-Graph   & 10.92$\pm$1.43 & 11.39 & 7.77--13.25 \\
ThingsBoard  & DKB (Ours) & 11.17$\pm$1.97 & 10.95 & 6.88--14.23 \\
ThingsBoard  & LLM-KB     & 15.29$\pm$4.94 & 13.17 & 9.62--26.46 \\
\midrule
OpenMRS Core & No-Graph   & 11.82$\pm$2.79 & 11.38 & 8.99--20.76 \\
OpenMRS Core & DKB (Ours) & 12.21$\pm$6.96 & 10.38 & 9.31--37.15 \\
OpenMRS Core & LLM-KB     & 11.94$\pm$4.88 & 11.09 & 8.12--27.81 \\
\bottomrule
\end{tabular}
\end{table*}

\subsection{Cost Analysis (Normalized)}
While absolute dollar costs depend on provider pricing, tokenization, model choice, and repository scale, \emph{relative} cost ratios provide a more stable comparison of cost drivers across retrieval strategies.
We therefore report \textbf{normalized end-to-end run cost} (indexing + answering the full 15-question suite), with the No-Graph baseline normalized to 1.0.
In our run, DKB incurred a modest cost multiple over No-Graph due to additional graph-aware context assembly, whereas LLM-KB exhibited a much larger multiplier dominated by \emph{LLM-mediated graph generation} during indexing (Table~\ref{tab:cost_abs})(Table~\ref{tab:cost_norm}).

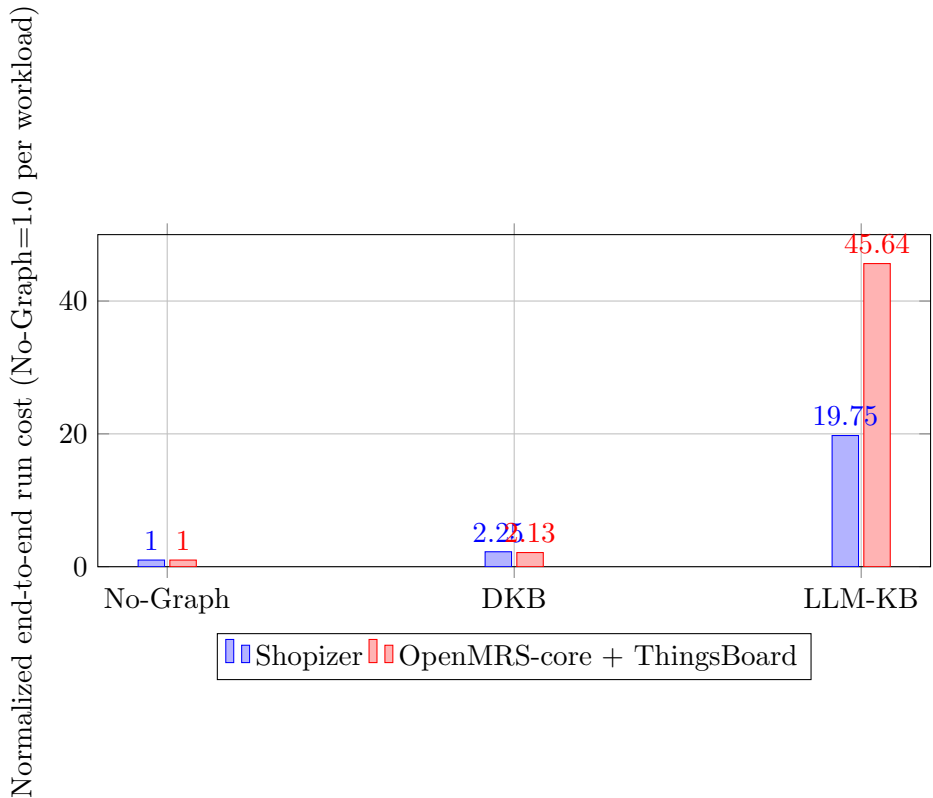
\begin{figure}[t]
\centering
\begin{tikzpicture}
\begin{axis}[
    ybar,
    bar width=10pt,
    ymin=0,
    ymax=50,
    ylabel={Normalized end-to-end run cost (No-Graph=1.0 per workload)},
    symbolic x coords={No-Graph,DKB,LLM-KB},
    xtick=data,
    legend style={at={(0.5,-0.20)},anchor=north,legend columns=2},
    nodes near coords,
    nodes near coords align={vertical},
    grid=major,
    width=\linewidth,
    height=6.0cm,
]
\addplot coordinates {(No-Graph,1.00) (DKB,2.25) (LLM-KB,19.75)};
\addplot coordinates {(No-Graph,1.00) (DKB,2.13) (LLM-KB,45.64)};
\legend{Shopizer, OpenMRS-core + ThingsBoard}
\end{axis}
\end{tikzpicture}
\caption{Normalized end-to-end run cost across workloads. Costs are normalized by the No-Graph baseline \emph{within each workload}. LLM-KB’s cost multiplier increases substantially on the larger workload, while DKB stays near $\sim$2$\times$.}
\label{fig:cost_all_workloads}
\end{figure}

To make this more portable beyond the Shopizer-scale repository, Figure~\ref{fig:cost_all_workloads} visualizes only the normalized ratios (No-Graph = 1.0), rather than treating absolute currency values as a generalizable claim.
In larger enterprise repositories, the same qualitative pattern is expected when LLM-KB performs extensive per-file or batched LLM extraction at index time: cost scales primarily with the volume of content processed by the model and any retries required for schema compliance.
This scaling pressure is often amplified in agentic RAG settings where retrieval, validation, and refinement may involve multiple tool calls and decision steps~\cite{Singh2025AgenticSurvey}.

\subsection{Cost Analysis on Multi-Repository Runs (OpenMRS + ThingsBoard)}
In addition to Shopizer, we measured end-to-end execution cost (indexing + answering all questions) for a combined run over \textit{two} repositories: OpenMRS-core and ThingsBoard.
Because absolute dollar costs depend on provider pricing and tokenization, we report both absolute totals for this combined run and normalized ratios with No-Graph set to 1.0.

\begin{table}[H]
\caption{End-to-end run cost (absolute USD) across workloads (indexing + questions).}
\label{tab:cost_abs}
\centering
\small
\setlength{\tabcolsep}{5pt}
\renewcommand{\arraystretch}{1.10}
\begin{tabular}{lccc}
\toprule
\textbf{Workload} & \textbf{No-Graph} & \textbf{DKB} & \textbf{LLM-KB} \\
\midrule
Shopizer (single repo, 15 questions) & \$0.04 & \$0.09 & \$0.79 \\
OpenMRS-core + ThingsBoard (15/repo = 30 total questions) & \$0.149 & \$0.317 & \$6.80 \\
\bottomrule
\end{tabular}
\end{table}

\begin{table}[H]
\caption{End-to-end run cost normalized to No-Graph (No-Graph = 1.0 per workload).}
\label{tab:cost_norm}
\centering
\small
\setlength{\tabcolsep}{5pt}
\renewcommand{\arraystretch}{1.10}
\begin{tabular}{lccc}
\toprule
\textbf{Workload} & \textbf{No-Graph} & \textbf{DKB} & \textbf{LLM-KB} \\
\midrule
Shopizer (single repo) & 1.00 & 2.25 & 19.75 \\
OpenMRS-core + ThingsBoard & 1.00 & 2.13 & 45.64 \\
\bottomrule
\end{tabular}
\end{table}

\noindent\textbf{Interpretation.}
The multi-repo cost pattern amplifies the same qualitative driver observed on Shopizer: LLM-KB’s total cost is dominated by LLM-mediated graph construction at indexing time, whereas DKB’s incremental cost over No-Graph is modest because its AST-derived graph is built deterministically and cheaply relative to LLM extraction.

\subsection{Indexing Reliability and Corpus Coverage}
Table~\ref{tab:coverage} summarizes corpus coverage signals taken directly from the three run logs.
All approaches discover the same number of Java files ($N_{\text{files}}=1210$), but diverge sharply in (i) how much code is embedded into the vector store and (ii) how complete the derived graph is.

\textbf{Visual contrast (graph artifacts).}
Beyond the numeric coverage counters, Fig.~\ref{fig:graphs} provides an immediate qualitative signal: the deterministic DKB artifact appears \emph{clustered/clean}, with coherent component-level groupings and fewer isolated fragments, whereas the LLM-KB artifact is comparatively \emph{fuzzy} and diffuse, consistent with extraction-time omissions (Table~\ref{tab:coverage}) and a less typed, schema-constrained dependency surface.

\begin{table*}[t]
\centering
\scriptsize
\caption{Code Coverage and Knowledge Extraction Completeness (Shopizer)}
\label{tab:coverage}
\begin{tabular}{lccc}
\toprule
\textbf{Metric} & \textbf{No-Graph} & \textbf{LLM-KB} & \textbf{DKB} \\
\midrule
Total Java files discovered & 1210 & 1210 & 1210 \\
Files skipped/missed by LLM & --- & 377 & --- \\
Files successfully analyzed & --- & 833 & --- \\
File success rate & --- & 0.688 & --- \\
\midrule
Total code chunks embedded & 5403 & 3465 & 4873 \\
Chunk coverage (vs No-Graph) & 1.000 & 0.641 & 0.902 \\
\midrule
Graph nodes & --- & 842 & 1158 \\
Graph edges & --- & 2552 & 1503 \\
Node coverage (vs DKB) & --- & 0.727 & 1.000 \\
\bottomrule
\end{tabular}
\end{table*}

\paragraph{Edge-count comparability.}
Raw edge counts are not directly comparable between DKB and LLM-KB because the two graphs encode different relation semantics and granularity.
DKB includes only a small, typed relation set (\texttt{injects}, \texttt{extends}, \texttt{implements}) derived from deterministic AST signals, whereas LLM-KB emits a broader \texttt{depends\_on} relation that can conflate usage, import-level references, and other non-inheritance dependencies.
As a result, LLM-KB can produce a denser edge set (more edges per node) even when its node set is smaller, and edge totals should be interpreted as \emph{schema-dependent} rather than as a pure measure of structural completeness.

\textbf{Observation 1 (Chunk shrinkage under LLM-KB).}
LLM-KB embeds 3465 chunks, substantially fewer than the vector-only baseline (5403) and fewer than deterministic DKB (4873).
This corresponds to \textbf{0.641 chunk coverage} relative to No-Graph, versus \textbf{0.902} for DKB (Table~\ref{tab:coverage}).
Because retrieval is constrained to indexed/embedded content, this shrinkage directly reduces the probability that evidence-bearing files can be retrieved at query time, especially for multi-hop questions whose supporting evidence is distributed across many files~\cite{Lewis2020Retrieval,Gao2023RetrievalSurvey}.

\textbf{Observation 2 (Probabilistic skipping + per-file success rate).}
The LLM-KB extraction log prints explicit ``\texttt{SKIPPED/MISSED by LLM}'' indicators, showing that the model’s batch-level structured output can omit some input files during indexing.
In the updated Shopizer run, the log flags \textbf{377} skipped/missed files, leaving \textbf{833} files successfully analyzed out of \textbf{1210} discovered, for a \textbf{per-file success rate of 0.688} (Table~\ref{tab:coverage}).
This highlights that Strategy~B is stochastic at the \emph{file} granularity (not only at the token or edge level): when a file is omitted from the structured output, it is also excluded from downstream embedding and graph insertion, creating systematic blind spots at query time.
While prompts, batching, and retries can reduce omissions, completeness remains mediated by stochastic model behavior and schema-bound extraction requirements, a known reliability risk in tool-mediated and agentic RAG pipelines~\cite{Singh2025AgenticSurvey}.

\textbf{Observation 3 (Graph size reduction and node coverage).}
Even after graph cleaning, LLM-KB produces a smaller graph (\textbf{842} nodes) than DKB (\textbf{1158} nodes), yielding \textbf{0.727 node coverage} (LLM-KB vs.\ DKB) (Table~\ref{tab:coverage}).
This reduction is consistent with extraction-time omissions and the resulting corpus shrinkage: because LLM-KB’s node set is derived from the model’s per-file structured records, skipped files can remove entire project-local entities from the graph and eliminate their edges, reducing the effectiveness of graph-based neighborhood expansion~\cite{OpenReview2025GraphSurvey}.

\paragraph{Cross-repository reliability summary.}
Across all three repositories, LLM-KB exhibits file-level extraction omission (File SR 0.650--0.806) that correlates with reduced chunk coverage (0.633--0.706), while DKB remains near baseline chunk coverage (0.902--0.993) without probabilistic file skipping (Tables~\ref{tab:coverage}, \ref{tab:thingsboard_coverage}, \ref{tab:openmrs_coverage}).

\textbf{Observation 4 (Why DKB chunk coverage is below the No-Graph baseline on Shopizer).}
Although DKB’s graph construction is deterministic, its embedding corpus is coupled to the set of Java files that successfully yield project-local type declarations and are mapped into the ontology (\texttt{file\_map}) during AST extraction.
In Shopizer, the No-Graph baseline embeds chunks from all discovered Java files, whereas DKB’s indexing path can exclude or reduce contributions from files that do not produce a parsable/mapped top-level type under the extractor (e.g., non-type utility files, atypical naming or path canonicalization mismatches, generated sources, or declarations that are not captured by the current Tree-sitter query patterns).
This explains why DKB’s chunk coverage on Shopizer (0.902 in Table~\ref{tab:coverage}) is slightly below the vector-only baseline even though DKB does not exhibit probabilistic file skipping.
Importantly, this gap reflects implementation coupling between graph node discovery and the embedding pipeline, not model stochasticity; decoupling embeddings from graph-node discovery (embed all discovered files, then attach graph edges where available) is a straightforward engineering improvement and would move DKB closer to the 0.99-level coverage observed on ThingsBoard and OpenMRS (Tables~\ref{tab:thingsboard_coverage}, \ref{tab:openmrs_coverage}). Future work will export a structured per-file embedding audit (file $\rightarrow$ chunk count) to quantify which Shopizer files contribute to the 9.8\% delta and verify that the effect is attributable to mapping/query coverage rather than hidden filtering. We treat this as an implementation-level coupling (and thus a fairness confound in chunk-count comparisons), not a limitation of AST-derived graph construction itself.

\subsection{Answer Correctness on Full 15-Question Suite}
Table~\ref{tab:correctness} summarizes correctness judgments for the full 15-question benchmark suite spanning repository semantics, multi-hop architectural tracing, and system-level reasoning.

\subsection{Generalization Across Repositories}
To test whether the observed trade-offs persist beyond Shopizer, we repeat the same three RAG variants on ThingsBoard and OpenMRS Core.
We report (i) indexing time (vector DB build and, where applicable, graph build), (ii) query latency over the 15-question suite, and (iii) answer correctness labels.

\subsubsection{ThingsBoard}
No-Graph indexes 4521 Java files into 32428 chunks \cite{tb_nograph_logs}.
DKB builds a Tree-sitter ontology with 4691 nodes and 8570 edges and indexes 32098 chunks \cite{tb_dkb_logs}.
LLM-KB builds a graph with 4299 nodes and 20419 edges and indexes 22899 chunks \cite{tb_llmkb_logs}.

\begin{table}[h]
\centering
\scriptsize
\caption{Code Coverage and Graph Completeness (ThingsBoard)}
\label{tab:thingsboard_coverage}
\begin{tabular}{lcccccc}
\toprule
\textbf{Method} & \textbf{Chunks} & \textbf{Chunk Cov.} & \textbf{Nodes} & \textbf{Edges} & \textbf{LLM-skipped} & \textbf{File SR} \\
\midrule
No-Graph & 32428 & 1.000 & --- & --- & --- & --- \\
LLM-KB & 22899 & 0.706 & 4299 & 20419 & 877 & 0.806 \\
DKB & 32098 & 0.990 & 4691 & 8570 & --- & --- \\
\bottomrule
\end{tabular}
\end{table}

\subsubsection{OpenMRS Core}
No-Graph indexes 1258 Java files into 11495 chunks \cite{openmrs_nograph_logs}.
DKB builds a Tree-sitter ontology with 1371 nodes and 1517 edges and indexes 11411 chunks \cite{openmrs_dkb_logs}.
LLM-KB builds a graph with 1052 nodes and 2957 edges and indexes 7281 chunks \cite{openmrs_llmkb_logs}.

\begin{table}[h]
\centering
\scriptsize
\caption{Code Coverage and Graph Completeness (OpenMRS-core)}
\label{tab:openmrs_coverage}
\begin{tabular}{lcccccc}
\toprule
\textbf{Method} & \textbf{Chunks} & \textbf{Chunk Cov.} & \textbf{Nodes} & \textbf{Edges} & \textbf{LLM-skipped} & \textbf{File SR} \\
\midrule
No-Graph & 11495 & 1.000 & --- & --- & --- & --- \\
LLM-KB & 7281 & 0.633 & 1052 & 2957 & 440 & 0.650 \\
DKB & 11411 & 0.993 & 1371 & 1517 & --- & --- \\
\bottomrule
\end{tabular}
\end{table}

\subsection{Correctness Summary}
To summarize Table~\ref{tab:correctness}, we collapse any ``Correct+'' detail level into \emph{Correct}.
Over the full 15-question Shopizer suite, DKB achieved the highest correctness (15/15), LLM-KB followed closely (13/15; 2 partial), while No-Graph lagged (6/15) and exhibited the highest hallucination risk on architecture-discovery queries (e.g., Q11).
This aligns with broader evidence that repository-level tasks stress retrieval adequacy and grounding across many files~\cite{Jimenez2023SWEbench,Tian2023RepoBench,Tao2025RetrievalSurvey}.

\begin{table}[H]
\caption{Correctness Summary on Full Shopizer Question Suite (from Table~\ref{tab:correctness})}
\label{tab:correctness_summary}
\centering
\begin{tabular}{lccc}
\toprule
\textbf{Approach} & \textbf{Correct} & \textbf{Partial} & \textbf{Incorrect} \\
\midrule
No-Graph & 6 & 4 & 5 \\
LLM-KB & 13 & 2 & 0 \\
DKB (Ours) & 15 & 0 & 0 \\
\bottomrule
\end{tabular}
\end{table}

\begin{table}[H]
\caption{Correctness Summary Across Repositories ($Q=15$ per repository; totals computed from Tables~\ref{tab:correctness}, \ref{tab:things_correctness}, \ref{tab:mrs_correctness}).}
\label{tab:correctness_summary_all}
\centering
\scriptsize
\setlength{\tabcolsep}{4.5pt}
\renewcommand{\arraystretch}{1.08}
\begin{tabular}{llccc}
\toprule
\textbf{Repository} & \textbf{Approach} & \textbf{Correct} & \textbf{Partial} & \textbf{Incorrect} \\
\midrule
Shopizer & No-Graph & 6 & 4 & 5 \\
Shopizer & LLM-KB & 13 & 2 & 0 \\
Shopizer & DKB (Ours) & 15 & 0 & 0 \\
\midrule
ThingsBoard & No-Graph & 14 & 1 & 0 \\
ThingsBoard & LLM-KB & 12 & 2 & 1 \\
ThingsBoard & DKB (Ours) & 14 & 1 & 0 \\
\midrule
OpenMRS Core & No-Graph & 11 & 4 & 0 \\
OpenMRS Core & LLM-KB & 13 & 1 & 1 \\
OpenMRS Core & DKB (Ours) & 14 & 1 & 0 \\
\midrule
\textbf{All (45 questions)} & No-Graph & 31 & 9 & 5 \\
\textbf{All (45 questions)} & LLM-KB & 38 & 5 & 2 \\
\textbf{All (45 questions)} & DKB (Ours) & 43 & 2 & 0 \\
\bottomrule
\end{tabular}
\end{table}

\subsection{Comparative Analysis}

\textbf{Deterministic Knowledge Base (DKB).}
Across the evaluated suites, DKB provides the most reliable structural grounding when questions require \emph{multi-hop reasoning} and \emph{upstream discovery} (e.g., identifying controllers/services that consume a dependency through interfaces).
This advantage comes from deterministic AST-derived relations that are frequently under-specified by text similarity alone (typed \texttt{injects}/\texttt{extends}/\texttt{implements} edges)~\cite{TreeSitter2025,Zhang2025CAST}.
In Shopizer, this is most visible on upstream questions such as controller discovery (Q11), where bidirectional traversal (successors and predecessors) materially improves evidence collection.
In particular, Strategy~C’s success on Q11 is driven by the \texttt{InterfaceConsumerExpand} rule in Alg.~\ref{alg:graph_rag}: when a retrieved concrete class implements an interface, the retriever additionally pulls in \emph{consumers of that interface} (predecessors of the interface node), allowing the system to cross interface boundaries and recover controllers/services that depend on the interface rather than a specific implementation.
We note that correctness gains are workload-dependent: on some suites (e.g., ThingsBoard) the vector-only baseline is already strong and DKB primarily matches it while preserving deterministic coverage and low indexing overhead (Table~\ref{tab:correctness_summary_all}).

\textbf{LLM-Generated Knowledge Base (LLM-KB).}
LLM-KB can achieve high correctness and produces useful high-level architectural summaries (e.g., identifying event-driven triggers such as Q9).
However, its indexing pipeline is probabilistic and schema-bound: run logs show that extraction can omit files (\texttt{SKIPPED/MISSED by LLM}), which reduces both the embedded corpus and the resulting graph footprint (Tables~\ref{tab:coverage}, \ref{tab:thingsboard_coverage}, \ref{tab:openmrs_coverage}).
This is not only a performance concern—missing files at index time create retrieval blind spots at query time.
LLM-KB also exhibits substantially higher end-to-end cost on larger workloads due to LLM-mediated graph construction (Tables~\ref{tab:cost_abs}, \ref{tab:cost_norm}).
Finally, raw edge totals should be interpreted cautiously: LLM-KB emits a broader \texttt{depends\_on} relation (often denser), while DKB edges are a narrower, typed subset (\texttt{injects}/\texttt{extends}/\texttt{implements}), so edge counts are not directly comparable as a completeness measure (Table~\ref{tab:graph_schema}).

\textbf{Vector-Only RAG (No-Graph).}
No-Graph performs well on localized questions when the necessary evidence is co-located within the top-$k$ retrieved chunks (e.g., Shopizer Q1 and Q15), and on some repositories it can be competitive overall (e.g., ThingsBoard in Table~\ref{tab:correctness_summary_all}).
Its main failure mode appears on questions where relevant evidence is distributed across multiple files connected by architectural structure (inheritance, dependency injection, interface-driven wiring).
When structural neighbors are not retrieved, the generator may fall back to framework conventions and produce ungrounded or partially grounded claims (e.g., Shopizer Q11), reflecting known risks when generation is weakly supported by retrieval~\cite{Lewis2020Retrieval,Gao2023RetrievalSurvey}.
Overall, No-Graph offers strong simplicity and often favorable latency, but it is less dependable for multi-hop architectural tracing unless augmented with structure-aware expansion.

% ------------------------
% Discussion
% ------------------------
\section{Discussion}
The results show a practical trade-off: LLM-generated indexing can be expensive and latency-unstable, while deterministic AST parsing produces a usable code graph at low indexing cost~\cite{TreeSitter2025,Edge2024GraphRAG}.
Although DKB is not universally faster at query time than vector-only retrieval, it substantially improves correctness on multi-hop structural questions by explicitly retrieving graph neighborhoods instead of relying on inferred relationships from disjoint text chunks~\cite{OpenReview2025GraphSurvey}.

\textbf{Why indexing reliability matters.}
Table~\ref{tab:coverage} shows that LLM-KB embeds substantially fewer chunks in comparison to the baseline, and its graph has fewer nodes than DKB.
This is an important distinction for enterprise settings: even if LLM-KB answers many questions correctly, omission of code artifacts at indexing time can create unpredictable blind spots.
Prompt improvements and retries may reduce skipping, but because the extraction is probabilistic and schema-bound, completeness cannot be guaranteed in multi-step/agentic workflows where intermediate structured outputs can fail validation~\cite{Singh2025AgenticSurvey}.

\textbf{Key takeaway:} For correctness-sensitive repository questions, deterministic structure provides more reliable grounding than probabilistic extraction, particularly for multi-hop and upstream discovery queries.

For correctness-sensitive tasks (impact analysis, controller discovery, transaction boundary inference), deterministic retrieval reduces hallucination by grounding the model in explicit topology and by preserving stable corpus coverage~.

% ------------------------
% Threats to Validity
% ------------------------
\section{Threats to Validity}
\subsection{Internal Validity}
Correctness labels were assigned using a human judgment protocol based on the repository’s ground truth.
Future work should include multiple annotators, inter-rater agreement, and explicit evidence citations per answer.
This aligns with broader concerns in evaluating repository-level systems where ground truth may be distributed and tool-mediated~\cite{Jimenez2023SWEbench,Tian2023RepoBench}.

\subsection{External Validity}
We evaluate on three Java repositories (Shopizer, ThingsBoard, and OpenMRS Core).
Results may differ for other architectures or languages.
Extending to multiple repos will strengthen generalization, consistent with repository-level benchmark findings that performance varies by repo topology and dependency structure~\cite{Tao2025RetrievalSurvey,Yang2025Empirical}.

\subsection{Construct Validity}
\textbf{Correctness construct.} The \emph{Correct/Partial/Incorrect} labels are intended to measure repository-grounded correctness: (i) correct entities (classes/methods/files), (ii) correct structural relationships (e.g., controller$\rightarrow$service$\rightarrow$repository chains, inheritance/DI edges), and (iii) avoidance of hallucinated components. These labels are necessarily coarse and may not capture nuance (e.g., an answer may be structurally correct but omit a key configuration detail).

\textbf{Stability across trials.} Although Table~\ref{tab:correctness} reports one representative run, we executed multiple trials for each pipeline under the same settings and observed similar correctness outcomes (and the same relative ordering of approaches). We do not treat this as a statistical guarantee; future work should report the exact number of trials, per-question agreement rate across trials, and confidence intervals.

\textbf{Coverage construct.} Coverage signals derived from run logs (e.g., discovered files, embedded chunk counts, graph node/edge counts, and explicit \texttt{SKIPPED/MISSED by LLM} indicators) are proxies for indexing completeness, but can be sensitive to instrumentation placement and path canonicalization. Future work should export these counters as structured artifacts (JSON/CSV), validate them against independent filesystem enumeration (stable unique file IDs), and add additional completeness checks (e.g., percentage of files producing at least one embedding; percentage of graph nodes mapped to an existing source file).

% ------------------------
% Reproducibility / Artifacts
% ------------------------
\section{Reproducibility and Artifact Availability}
\label{sec:reproducibility}
All experimental code and (sanitized) run outputs for this paper will be released in the companion repository:
\url{https://github.com/Manideep-Reddy-Chinthareddy/graph-based-rag-ast-vs-llm}.
To ensure bitwise-identical retrieval results are feasible, we recommend pinning (i) a repository commit hash for both the evaluation code and the target codebase, and (ii) the exact Python dependency versions used to build embeddings and graphs.

\subsection{Artifacts Provided}
The repository contains three end-to-end evaluation scripts corresponding to the compared pipelines:
(i) \texttt{nograph\_gemini.py} (vector-only Naive RAG),
(ii) \texttt{llmkb\_gemini.py} (LLM-generated knowledge graph RAG),
and (iii) \texttt{dkb\_gemini\_v2.py} (deterministic Tree-sitter AST graph RAG).
Each script (a) scans the target repository, (b) builds the index (and graph when applicable), (c) executes the same 15-question benchmark suite, and (d) writes a structured JSON file with per-question answers and latency measurements.

\subsection{Target Codebase and Version Pinning}
Our experiments use a pinned fork of the Shopizer Java repository to keep the target codebase static during analysis:
\url{https://github.com/Manideep-Reddy-Chinthareddy/shopizer-fork}. \cite{ShopizerRepo}
For strict reproducibility, the exact fork commit used in evaluation should be recorded and reported as:
\texttt{SHOPIZER\_FORK\_COMMIT=6a4a0a65a3408ee8f62\\597b51d1b3aac24b77dee}.
Likewise, the artifact repository commit should be recorded as:
\texttt{ARTIFACT\_COMMIT=efa7c88ff998877d82\\5b722a595b9f0201719da4}.
All reported counts (e.g., discovered Java files, embedded chunk counts, graph node/edge counts) are a function of these pinned commits.

\subsection{Environment and Dependencies}
All pipelines are implemented in Python and rely on:
(1) a vector store for embedding retrieval,
(2) an embedding model and LLM provider used through a single framework abstraction,
and (3) (for DKB) an incremental parsing stack based on Tree-sitter~\cite{TreeSitter2025}.
To reproduce results, create a clean virtual environment and install the exact pinned dependencies from \texttt{requirements.txt} (recommended) or an equivalent lockfile.
The scripts expect the LLM provider API key to be available via environment variables (e.g., \texttt{GOOGLE\_API\_KEY} when using Gemini through LangChain wrappers).

\subsection{Running the Benchmarks}
All three scripts are designed to be executed as standalone entry points.
The only required local configuration is setting the target repository path (the Shopizer clone) in the script configuration (e.g., \texttt{PROJECT\_ROOT}) or via an environment variable if the repository exposes one.
Each script runs the same fixed suite of $Q=15$ benchmark questions (embedded directly in the script) using shared retrieval hyperparameters:
\texttt{chunk\_size=1000}, \texttt{chunk\_overlap=100}, and top-$k=10$ vector retrieval.

\subsection{Outputs, Logs, and Metric Extraction}
Each pipeline writes a JSON artifact containing:
(i) indexing time (\texttt{db\_gen\_time}),
(ii) graph build time when applicable (\texttt{graph\_generation\_time}),
and (iii) a list of per-question responses with end-to-end latency (\texttt{time\_taken\_seconds}).
Default output filenames are:
\texttt{nograph\_gemini\_rag\_response.json},
\texttt{llmkb\_gemini\_rag\_response.json},
and \texttt{dkb\_gemini\_v2\_rag\_response.json}.
For graph-based methods, the scripts additionally emit rendered graph visualizations (HTML) and/or intermediate logs for coverage auditing (e.g., explicit \ttfamily SKIPPED/MISSED by LLM \rmfamily\ indicators for LLM-KB).

\subsection{Reproducing Summary Tables}
Table~\ref{table:metrics} is computed directly from the per-question latency fields in the JSON artifacts (mean, standard deviation, median, and min--max).
Table~\ref{tab:coverage} is computed from the indexing logs and JSON counters emitted by each pipeline (discovered files, embedded chunks, and graph size).
To reproduce the reported numbers, users should:
(1) run all three scripts on the same pinned Shopizer commit,
(2) keep retrieval hyperparameters identical across runs,
and (3) derive aggregate statistics from the emitted JSONs without manual editing.

\subsection{Notes on Stochasticity}
Although DKB’s graph construction is deterministic (AST-derived) given a fixed codebase and parser version, LLM-KB’s extraction is inherently stochastic due to model variability and schema compliance failure modes, and because non-greedy sampling schemes can produce a distribution of outputs even for identical prompts. ~\cite{Singh2025AgenticSurvey,Hayes2024ProbExtract}.
To quantify variance, we recommend repeating LLM-KB indexing and evaluation for multiple trials under identical settings, and reporting the distribution of (i) skipped-file counts, (ii) embedded chunk counts, and (iii) latency outliers.

% ------------------------
% Ethical Considerations
% ------------------------
\section{Ethical Considerations}
Graph-RAG for codebases can expose sensitive implementation details if used on proprietary repositories.
Enterprise deployments should enforce access control, auditing, and avoid transmitting restricted code to third-party services without appropriate approval.
Agentic and tool-using patterns can amplify both capability and risk in software engineering deployments~\cite{Singh2025AgenticSurvey,GitHub2025SoftwareDev}.

% ------------------------
% Limitations and Future Work
% ------------------------
\section{Limitations and Future Work}
Deterministic AST extraction can miss behaviors driven by reflection, runtime-generated code, or dynamic dispatch that is not explicit in syntax~\cite{TreeSitter2025}.
Future work includes hybrid graphs (deterministic backbone + LLM semantic edges), richer call graph resolution, and evaluation across multiple languages~\cite{OpenReview2025GraphSurvey,Tao2025RetrievalSurvey}.

For LLM-KB, future work should (i) export explicit counters for skipped files and schema failures, (ii) add retries with smaller batch sizes, (iii) add static validation passes that ensure every discovered Java file yields a corresponding analyzed record, and (iv) decouple document embedding from LLM-derived class maps to prevent corpus shrinkage when graph extraction is incomplete~\cite{Singh2025AgenticSurvey}.

% ------------------------
% Conclusion
% ------------------------
\section{Conclusion}
We benchmarked vector-only RAG, an LLM-generated graph RAG (LLM-KB), and a deterministic AST-derived graph RAG (DKB) for repository-level code analysis on Shopizer, with additional runs on ThingsBoard and OpenMRS Core~\cite{Lewis2020Retrieval,Edge2024GraphRAG,TreeSitter2025}.
Across repositories, DKB consistently delivered strong correctness with low and predictable indexing overhead: on Shopizer, DKB achieved 15/15 correct answers while remaining close to the vector baseline in build time (22.09s vs.\ 18.41s), whereas LLM-KB incurred large offline graph-generation overhead (215.09s total) despite competitive query-time performance.

A key finding is that LLM-KB’s indexing pipeline can be \emph{incomplete} at the file level.
On Shopizer, the run logs report 377 files \texttt{SKIPPED/MISSED by LLM}, yielding a per-file success rate of 0.688 (833/1210).
This incompleteness correlates with reduced embedded corpus size: LLM-KB indexed 3465 chunks (0.641 coverage vs.\ the No-Graph baseline of 5403), while DKB indexed 4873 chunks (0.902 coverage).
The extracted graph footprint is also smaller under LLM-KB on Shopizer (842 nodes) compared to DKB (1158 nodes), which can reduce the effectiveness of graph neighborhood expansion for multi-hop queries.

Cost measurements further highlight the trade-off.
On Shopizer, end-to-end cost (indexing + 15 questions) was \$0.04 for No-Graph, \$0.09 for DKB, and \$0.79 for LLM-KB; on the larger combined OpenMRS-core + ThingsBoard workload, costs were \$0.149 (No-Graph), \$0.317 (DKB), and \$6.80 (LLM-KB), widening the normalized multiplier for LLM-KB substantially.
Overall, when correctness and coverage reliability matter, DKB provides a practical ``compiler-in-the-loop'' retrieval strategy: it preserves indexing completeness signals close to the vector baseline while improving multi-hop architectural grounding without the high and variable cost of LLM-mediated graph construction.

Overall, the cross-repository pattern is consistent: DKB provides a low-overhead, deterministic structural signal and near-baseline coverage, while LLM-KB can achieve strong correctness with lesser code to implement but introduces indexing-time incompleteness and cost inflation due to probabilistic, schema-bound extraction.

\appendix

\section{Per-question correctness tables}
This section shows the full detailed view of the questions and correctness for each of the question per repository/RAG model
\subsection{Shopizer}
\begingroup
\setlength{\tabcolsep}{3.2pt}
\renewcommand{\arraystretch}{1.06}
\scriptsize

\begin{longtable}{p{0.11\textwidth} p{0.27\textwidth} p{0.18\textwidth} p{0.18\textwidth} p{0.18\textwidth}}
\caption{Correctness Comparison of RAG Strategies on the Full Shopizer Question Suite ($Q=15$)}
\label{tab:correctness}\\
\toprule
\textbf{Question} & \textbf{Theme / Ground-Truth Signal} & \textbf{No-Graph (Vector Only)} & \textbf{LLM-KB (Probabilistic)} & \textbf{DKB (Deterministic)} \\
\midrule
\endfirsthead

\multicolumn{5}{l}{\small \textbf{Table \thetable\ (continued)}}\\
\toprule
\textbf{Question} & \textbf{Theme / Ground-Truth Signal} & \textbf{No-Graph (Vector Only)} & \textbf{LLM-KB (Probabilistic)} & \textbf{DKB (Deterministic)} \\
\midrule
\endhead

\bottomrule
\endlastfoot
Q1 & Repository save implementation & 
\textbf{Correct} (Identified SimpleJpaRepository, proxy pattern.) &
\textbf{Correct} (Identified proxy interceptors, JPA repository delegation.) &
\textbf{Correct} (Identified SimpleJpaRepository, proxy pattern.) \\
\midrule
Q2 & UserStoreHandler responsibilities &
\textbf{Correct} (Identified user store operations and related handlers.) &
\textbf{Correct} (Identified UserStoreHandler responsibilities and operations.) &
\textbf{Correct} (Identified handler responsibilities, authentication integration.) \\
\midrule
Q3 & Custom vs generic methods &
\textbf{Incorrect} (Assumed no custom methods, missed overrides.) &
\textbf{Partial} (Mentioned custom methods but missed some key overrides.) &
\textbf{Correct} (Identified custom methods and overrides accurately.) \\
\midrule
Q4 & PricingService logic &
\textbf{Partial} (Gave general pricing description, missed ProductPriceUtils.) &
\textbf{Correct} (Identified ProductPriceUtils and pricing flow.) &
\textbf{Correct} (Identified pricing calculation details and ProductPriceUtils.) \\
\midrule
Q5 & Inventory updates and triggers &
\textbf{Incorrect} (Hallucinated direct inventory update flow.) &
\textbf{Correct} (Identified inventory updates and relevant services.) &
\textbf{Correct} (Identified triggers and correct inventory update path.) \\
\midrule
Q6 & CustomerService.save flow &
\textbf{Partial} (Gave generic repository save flow, missed key chain.) &
\textbf{Correct} (Traced correct service chain and region resolution.) &
\textbf{Correct} (Identified correct inheritance chain and resolution.) \\
\midrule
Q7 & Checkout flow trace &
\textbf{Partial} (Identified entry but missed downstream services.) &
\textbf{Correct} (Identified controllers and key downstream services.) &
\textbf{Correct} (Identified auth checks, payment/shipping services.) \\
\midrule
Q8 & Authentication / authorization handling &
\textbf{Correct} (Identified auth utilities and flow.) &
\textbf{Correct} (Identified auth flow and relevant security components.) &
\textbf{Correct} (Identified auth flow and relevant security components.) \\
\midrule
Q9 & SearchService trigger &
\textbf{Incorrect} (Hallucinated direct trigger instead of event-driven.) &
\textbf{Correct} (Identified event-driven trigger for SearchService.) &
\textbf{Correct} (Identified IndexProductEventListener / event flow.) \\
\midrule
Q10 & Impact of deletion (module breakage) &
\textbf{Partial} (Gave generic cascade impact.) &
\textbf{Correct} (Identified impacted payment/shipping integrations.) &
\textbf{Correct} (Identified Stripe/USPS/Facebook API integration impact.) \\
\midrule
Q11 & Controllers using Shopping Cart logic &
\textbf{Incorrect} (Hallucinated controllers and missed interface usage.) &
\textbf{Correct} (Identified OrderApi and related controllers.) &
\textbf{Correct} (Identified interface-consumer expansion across boundary.) \\
\midrule
Q12 & Product / category relationship logic &
\textbf{Correct} (Identified relationship mappings and relevant entities.) &
\textbf{Correct} (Identified mappings and relationship logic.) &
\textbf{Correct} (Identified relationship mappings and correct logic.) \\
\midrule
Q13 & Transactional proof (createOrder multi-write boundary) &
\textbf{Incorrect} (Missed transactional annotations and evidence.) &
\textbf{Partial} (Identified unit-of-work but missed some boundaries.) &
\textbf{Correct} (Identified transactional scope across multi-repo writes.) \\
\midrule
Q14 & Shipping module logic & 
\textbf{Correct} (Identified shipping integrations and service flow.) &
\textbf{Correct} (Identified shipping module and service logic.) &
\textbf{Correct} (Identified shipping module and service logic.) \\
\midrule
Q15 & Order ID generation &
\textbf{Correct} (Identified @TableGenerator and table-based sequencing.) &
\textbf{Correct} (Identified SM\_SEQUENCER and table generator usage.) &
\textbf{Correct} (Explained atomic increment and sequencing table usage.) \\
\bottomrule
\end{longtable}
\endgroup
\subsection{ThingsBoard}
% ------------------------
% ThingsBoard: Correctness table (Shopizer Table-3 format)
% ------------------------
\begingroup
\setlength{\tabcolsep}{3.2pt}
\renewcommand{\arraystretch}{1.06}
\scriptsize

\begin{longtable}{p{0.11\textwidth} p{0.27\textwidth} p{0.18\textwidth} p{0.18\textwidth} p{0.18\textwidth}}
\caption{Correctness Comparison of RAG Strategies on the ThingsBoard Question Suite ($Q=15$)}
\label{tab:things_correctness}\\
\toprule
\textbf{Question} & \textbf{Theme / Ground-Truth Signal} & \textbf{No-Graph (Vector Only)} & \textbf{LLM-KB (Probabilistic)} & \textbf{DKB (Deterministic)} \\
\midrule
\endfirsthead

\multicolumn{5}{l}{\small \textbf{Table \thetable\ (continued)}}\\
\toprule
\textbf{Question} & \textbf{Theme / Ground-Truth Signal} & \textbf{No-Graph (Vector Only)} & \textbf{LLM-KB (Probabilistic)} & \textbf{DKB (Deterministic)} \\
\midrule
\endhead

\bottomrule
\endlastfoot
Q1 & App entry point \& startup & 
\textbf{Correct} (Identified \texttt{ThingsboardServerApplication} / Spring Boot startup.) &
\textbf{Incorrect} (Claimed no single entry point; missed \texttt{ThingsboardServerApplication}.) &
\textbf{Correct} (Identified \texttt{ThingsboardServerApplication} and initial bootstrap.) \\
\midrule
Q2 & Login/authentication path &
\textbf{Correct} (Traced \texttt{AuthController} \textrightarrow{} auth manager/provider \textrightarrow{} token response.) &
\textbf{Correct} (Identified security filter/provider chain and JWT/token creation.) &
\textbf{Correct} (Traced request \textrightarrow{} authentication \textrightarrow{} token issuance clearly.) \\
\midrule
Q3 & Upstream consumers of auth &
\textbf{Correct} (Named dependent controllers/filters using the same token/auth components.) &
\textbf{Correct} (Identified shared auth filters/components and their consumers.) &
\textbf{Correct} (Found upstream consumers of auth utilities across controllers/filters.) \\
\midrule
Q4 & Telemetry ingestion pipeline &
\textbf{Partial} (High-level service chain; missing a fully grounded end-to-end trace.) &
\textbf{Partial} (Described generic TB telemetry path; incomplete concrete call chain.) &
\textbf{Partial} (Explained ingestion/actor handoff but noted missing controller evidence.) \\
\midrule
Q5 & MQTT transport handler flow &
\textbf{Correct} (Traced MQTT inbound decode \textrightarrow{} message creation \textrightarrow{} next dispatch.) &
\textbf{Correct} (Identified message object + next-hop dispatch in transport layer.) &
\textbf{Correct} (Traced handler \textrightarrow{} message conversion \textrightarrow{} routing correctly.) \\
\midrule
Q6 & Netty pipeline initialization &
\textbf{Correct} (Listed pipeline handlers and responsibilities in initializer.) &
\textbf{Correct} (Recovered handler order and roles consistent with Netty init.) &
\textbf{Correct} (Identified initializer wiring: codec/SSL/handlers in order.) \\
\midrule
Q7 & TbContext responsibilities &
\textbf{Correct} (Explained \texttt{TbContext} as runtime services context for rule nodes.) &
\textbf{Correct} (Described \texttt{TbContext} role + service access for rule nodes.) &
\textbf{Correct} (Identified \texttt{TbContext} responsibilities and providers.) \\
\midrule
Q8 & RuleEngine tellNext message &
\textbf{Correct} (Explained \texttt{RuleNodeToRuleChainTellNextMsg} next-hop routing.) &
\textbf{Correct} (Identified tellNext as rule-chain transition message; named consumer.) &
\textbf{Correct} (Explained actor consumption + next-hop logic for tellNext.) \\
\midrule
Q9 & Alarm API to persistence &
\textbf{Correct} (Traced AlarmController \textrightarrow{} service \textrightarrow{} DAO/repository persistence.) &
\textbf{Correct} (Identified controller/service/DAO chain down to persistence.) &
\textbf{Correct} (Correct end-to-end alarm flow to persistence.) \\
\midrule
Q10 & Dashboard access control &
\textbf{Correct} (Located permission/tenant checks and enforcing classes.) &
\textbf{Correct} (Identified access-control checks at controller/service boundaries.) &
\textbf{Correct} (Pinned enforcement points for dashboard permissions.) \\
\midrule
Q11 & YAML config load/override &
\textbf{Correct} (Explained Spring config-name/properties binding; noted \texttt{tb-mqtt-transport.yml}.) &
\textbf{Partial} (Focused on test config props; did not pin runtime binding/override path.) &
\textbf{Correct} (Identified runtime config binding/override mechanism for transport YAML.) \\
\midrule
Q12 & Timeseries storage backend &
\textbf{Correct} (Identified timeseries DAO/repository layer and backend selection logic.) &
\textbf{Correct} (Named primary timeseries persistence components and selection.) &
\textbf{Correct} (Correct timeseries storage path and main DAO classes.) \\
\midrule
Q13 & Queue type selection &
\textbf{Correct} (Pointed to config-driven Kafka/Rabbit/etc selection and ownership.) &
\textbf{Correct} (Identified where queue impl is chosen and producer/consumer modules.) &
\textbf{Correct} (Recovered config points selecting queue implementation.) \\
\midrule
Q14 & ActorSystemContext + routing &
\textbf{Correct} (Explained ActorSystemContext construction and tenant/device partition routing.) &
\textbf{Correct} (Identified routing/partition providers for tenant/device decisions.) &
\textbf{Correct} (Traced construction + routing components for partitioning.) \\
\midrule
Q15 & Add transport or RuleNode &
\textbf{Correct} (Outlined required interfaces + registration/discovery locations.) &
\textbf{Correct} (Explained extension points and registration mechanism.) &
\textbf{Correct} (Gave correct steps to add new transport or rule node.) \\
\bottomrule
\end{longtable}
\endgroup
\subsection{OpenMRS Core}
% ------------------------
% OpenMRS-core: Correctness table (Shopizer Table-3 format)
% ------------------------
\begingroup
\setlength{\tabcolsep}{3.2pt}
\renewcommand{\arraystretch}{1.06}
\scriptsize

\begin{longtable}{p{0.11\textwidth} p{0.27\textwidth} p{0.18\textwidth} p{0.18\textwidth} p{0.18\textwidth}}
\caption{Correctness Comparison of RAG Strategies on the OpenMRS Question Suite ($Q=15$)}
\label{tab:mrs_correctness}\\
\toprule
\textbf{Question} & \textbf{Theme / Ground-Truth Signal} & \textbf{No-Graph (Vector Only)} & \textbf{LLM-KB (Probabilistic)} & \textbf{DKB (Deterministic)} \\
\midrule
\endfirsthead

\multicolumn{5}{l}{\small \textbf{Table \thetable\ (continued)}}\\
\toprule
\textbf{Question} & \textbf{Theme / Ground-Truth Signal} & \textbf{No-Graph (Vector Only)} & \textbf{LLM-KB (Probabilistic)} & \textbf{DKB (Deterministic)} \\
\midrule
\endhead

\bottomrule
\endlastfoot
Q1 & Web init + Context session lifecycle & 
\textbf{Correct} (Identified \texttt{Listener} + \texttt{OpenmrsFilter} session open/close.) &
\textbf{Correct} (Traced Listener/bootstrap and request filter managing Context sessions.) &
\textbf{Correct} (Identified \texttt{Listener}/\texttt{OpenmrsFilter} order-of-calls.) \\
\midrule
Q2 & Service lookup / ServiceContext wiring &
\textbf{Correct} (Explained \texttt{Context.getService} via Spring-backed \texttt{ServiceContext}.) &
\textbf{Correct} (Identified \texttt{ServiceContext} as Spring registry for service beans.) &
\textbf{Correct} (Traced Context facade \textrightarrow{} \texttt{ServiceContext} \textrightarrow{} service impl.) \\
\midrule
Q3 & Authentication path &
\textbf{Correct} (Traced \texttt{Context.authenticate} \textrightarrow{} user service check \textrightarrow{} UserContext.) &
\textbf{Correct} (Identified credential validation + storing authenticated user in context.) &
\textbf{Correct} (Correct end-to-end auth flow and where user is stored.) \\
\midrule
Q4 & Privilege enforcement &
\textbf{Correct} (Identified \texttt{@Authorized}/AOP advice + requirePrivilege checks.) &
\textbf{Correct} (Located privilege checks in AOP/authorization layer and call sites.) &
\textbf{Correct} (Correctly described authorization advice + upstream callers.) \\
\midrule
Q5 & saveEncounter chain + transaction boundary &
\textbf{Partial} (Traced service \textrightarrow{} DAO chain but did not pin \texttt{@Transactional} boundary.) &
\textbf{Correct} (Located \texttt{@Transactional} at service layer around saveEncounter.) &
\textbf{Correct} (Identified service-layer transaction boundary and DAO persistence.) \\
\midrule
Q6 & Upstream consumers of saveEncounter &
\textbf{Partial} (Gave generic HL7/import consumers without concrete caller classes.) &
\textbf{Incorrect} (Listed lifecycle/handler concepts; missing concrete upstream callers of saveEncounter.) &
\textbf{Correct} (Named concrete upstream consumers like \texttt{ORUR01Handler} and builders.) \\
\midrule
Q7 & VisitService.saveVisit child handling &
\textbf{Partial} (Discussed encounter/visit handlers but not \texttt{VisitServiceImpl.saveVisit} specifics.) &
\textbf{Partial} (High-level description; lacks grounded saveVisit/Obs persistence behavior.) &
\textbf{Partial} (Explained related handler semantics; missing direct saveVisit implementation proof.) \\
\midrule
Q8 & Obs \textrightarrow{} Concept + validation &
\textbf{Correct} (Explained Concept linkage + datatype/numeric validation points.) &
\textbf{Correct} (Identified Concept references and where Obs value validation occurs.) &
\textbf{Correct} (Correctly described validation + concept binding.) \\
\midrule
Q9 & Global properties cache/store &
\textbf{Correct} (Traced AdministrationService global property persistence + caching.) &
\textbf{Correct} (Identified storage, cache, and retrieval path for global properties.) &
\textbf{Correct} (Correct end-to-end global property flow.) \\
\midrule
Q10 & User create/update persistence &
\textbf{Correct} (Traced \texttt{UserServiceImpl.saveUser} \textrightarrow{} DAO \textrightarrow{} domain objects.) &
\textbf{Correct} (Identified service-to-DAO flow, including credentials/person linkage.) &
\textbf{Correct} (Correct user save flow down to persistence.) \\
\midrule
Q11 & Module classloading &
\textbf{Correct} (Explained \texttt{OpenmrsClassLoader} vs \texttt{ModuleClassLoader} responsibilities.) &
\textbf{Correct} (Identified module load/start classloading usage sites.) &
\textbf{Correct} (Correct mapping of classloader roles and usage.) \\
\midrule
Q12 & ModuleActivator lifecycle &
\textbf{Correct} (Located activator invocation path during module startup/shutdown.) &
\textbf{Correct} (Identified ModuleFactory invoking activators and call path.) &
\textbf{Correct} (Correct lifecycle call path for activators.) \\
\midrule
Q13 & Daemon elevated execution &
\textbf{Correct} (Explained daemon thread execution and privilege/user context for daemon runs.) &
\textbf{Correct} (Identified daemon token/elevation pattern and context setup.) &
\textbf{Correct} (Correct daemon mechanism and context/privileges.) \\
\midrule
Q14 & messages.properties + locale &
\textbf{Correct} (Traced MessageSource lookup and where locale is chosen.) &
\textbf{Correct} (Identified message resolution + locale selection path.) &
\textbf{Correct} (Correct i18n message resolution flow.) \\
\midrule
Q15 & Core flow + wiring locations &
\textbf{Partial} (Reasonable layered flow but did not name concrete Spring wiring artifacts.) &
\textbf{Correct} (Identified module/service boundaries and Spring wiring locations at a high level.) &
\textbf{Correct} (Mapped boundaries and where wiring is defined/loaded.) \\
\bottomrule
\end{longtable}
\endgroup

\section{Prompts}
This appendix lists the exact prompt templates used by each pipeline implementation.

\subsection{No-Graph (Vector-only) RAG: Answer Prompt}
\begin{lstlisting}[caption={Answer prompt used by the No-Graph (vector-only) baseline.}, label={lst:prompt_nograph}]
You are a Senior Java Engineer. Answer based on the code provided.

Context:
{context}

Question: {question}
\end{lstlisting}

\subsection{DKB (Tree-sitter AST Graph RAG): Answer Prompt}
\begin{lstlisting}[caption={Answer prompt used by DKB (AST-derived graph) with ontology-aware context.}, label={lst:prompt_dkb}]
You are a Senior Java Engineer. Answer based on the code provided.
Use the [ONTOLOGY INFO] to understand the architecture.

Context:
{context}

Question: {question}
\end{lstlisting}

\subsection{LLM-KB (LLM-Generated Graph): Index-Time Extraction Prompt}
\begin{lstlisting}[caption={Extraction prompt used to produce per-file class + dependency records for LLM-KB graph construction.}, label={lst:prompt_llmkb_extract}]
You are a static code analyzer. Analyze the provided batch of Java files.
For EACH file, identify the Class Name and any Custom Dependencies (other classes in the project that are used).
Ignore standard libraries (java.*, spring framework, etc.).

Return a JSON object with a single key "results" containing a list of objects, one for each file.

Schema:
{
  "results": [
    {
      "file_path": "path/to/File.java",
      "class_name": "NameOfClass",
      "dependencies": ["OtherClass1", "OtherClass2"]
    }
  ]
}

Input Files:
{batch_content}
\end{lstlisting}

\subsection{LLM-KB (LLM-Generated Graph): Answer Prompt}
\begin{lstlisting}[caption={Answer prompt used by LLM-KB at query time (vector hits + graph-expanded neighbors).}, label={lst:prompt_llmkb_answer}]
You are a Senior Java Architect. Answer the question based on the provided Code Context.
The context includes the main classes found via vector search, AND their dependencies from the Gemini-generated Knowledge Graph.

Context:
{context}

Question: {question}
\end{lstlisting}

\section{Graph Schema and Edge Types (Implemented)}
This appendix specifies the concrete node/edge schema and query-time expansion rules used by the two graph-based systems in this paper: the deterministic Tree-sitter graph (DKB) and the LLM-extracted graph (LLM-KB).

\subsection{DKB (Tree-sitter) Ontology Graph}
DKB builds a directed graph $G=(V,E)$ by parsing Java source files using Tree-sitter queries for type discovery and dependency signals (field types and constructor parameter types). Each discovered type is mapped to its defining file path and inserted as a node with a \texttt{type} attribute (\texttt{class}, \texttt{interface}, \texttt{enum}, \texttt{record}, \texttt{annotation}) and a \texttt{path} attribute. This node typing is derived directly from the Tree-sitter declaration node kind (\texttt{*\_declaration}) at indexing time.

\subsubsection{Node Types}
Nodes represent project-local Java types discovered by parsing declarations:
\begin{itemize}
  \item \textbf{class} (\texttt{class\_declaration})
  \item \textbf{interface} (\texttt{interface\_declaration})
  \item \textbf{enum} (\texttt{enum\_declaration})
  \item \textbf{record} (\texttt{record\_declaration})
  \item \textbf{annotation} (\texttt{annotation\_type\_declaration})
\end{itemize}

\subsubsection{Edge Types (Relation Labels)}
Edges are directed from a \emph{source type} to a \emph{target type} with a \texttt{relation} label:
\begin{itemize}
  \item \textbf{\texttt{injects}}: emitted when a class has a field whose declared \texttt{type\_identifier} matches another project type, or when a constructor formal parameter type matches another project type. This is a lightweight proxy for dependency injection and composition.
  \item \textbf{\texttt{extends}}: emitted when the Tree-sitter AST provides a \texttt{superclass} (classes) or extended \texttt{interfaces} (interfaces).
  \item \textbf{\texttt{implements}}: emitted when a class/record/enum declares \texttt{interfaces}.
\end{itemize}

\subsubsection{Inheritance Extraction Rules}
Inheritance relationships are extracted by reading the declaration node fields (e.g., \texttt{superclass} and \texttt{interfaces}) and mapping each \texttt{type\_identifier} in those fields to a known project type. For interface declarations, \texttt{interfaces} is treated as an \texttt{extends} relation (interfaces extending interfaces).

\subsubsection{Query-time Expansion Semantics (Bidirectional + Interface-Consumer Expansion)}
At query time, DKB performs:
\begin{enumerate}
  \item \textbf{Initial retrieval:} top-$k$ vector retrieval over code chunks.
  \item \textbf{Bidirectional neighborhood expansion:} for each retrieved class $v$, include both successors $\textsc{Succ}(v)$ (downstream dependencies) and predecessors $\textsc{Pred}(v)$ (upstream consumers).
  \item \textbf{Interface-consumer expansion (critical fix):} if a retrieved concrete class $c$ has an \texttt{implements} edge to an interface $I$, then additionally include \emph{predecessors of $I$} (classes that inject/use the interface), improving upstream controller discovery across interface boundaries.
\end{enumerate}

\subsubsection{Context Assembly Budgeting}
To bound prompt size, DKB emits (i) an excerpt of each initially retrieved chunk and (ii) bounded excerpts from neighbor files when expanding successors and predecessors, while also emitting a lightweight relationship summary (\texttt{[ONTOLOGY INFO]}) to make the traversal trace visible.

\subsection{LLM-KB (LLM-extracted) Dependency Graph}
LLM-KB builds a directed graph by batching Java files into prompts and requiring the model to return a structured list of records: (\texttt{file\_path}, \texttt{class\_name}, \texttt{dependencies}). Nodes are added for each extracted \texttt{class\_name}, and edges are added from the class to each dependency with relation label \texttt{depends\_on}. Non-project dependencies are intended to be excluded by instruction (e.g., \texttt{java.*}, Spring framework packages).

\subsubsection{Node Types}
LLM-KB treats all extracted entities as:
\begin{itemize}
  \item \textbf{class}: nodes added from LLM-emitted \texttt{class\_name}.
\end{itemize}

\subsubsection{Edge Types}
\begin{itemize}
  \item \textbf{\texttt{depends\_on}}: directed edge $c \rightarrow d$ for each dependency $d$ emitted by the LLM for class $c$.
\end{itemize}

\subsubsection{Indexing Completeness Signals}
Because extraction is schema-bound and performed in batches under context constraints, the implementation truncates file content per prompt batch and explicitly detects when an input file does not appear in the structured output, emitting \texttt{SKIPPED/MISSED by LLM} for missing files. This signal is used in the paper as direct evidence of probabilistic extraction incompleteness.

\subsubsection{Query-time Expansion Semantics}
At query time, LLM-KB performs top-$k$ vector retrieval and then expands to graph neighbors using:
\begin{itemize}
  \item \textbf{Successors:} dependencies of the retrieved class (downstream).
  \item \textbf{Predecessors:} classes that reference the retrieved class (upstream).
\end{itemize}
Unlike DKB, LLM-KB does not enforce interface-aware expansion based on explicit \texttt{implements}/\texttt{extends} edges, because the extracted schema does not distinguish inheritance from usage.

\subsection{Summary: Why the Schema Matters}
The DKB schema is \emph{typed} (distinguishing \texttt{extends} vs.\ \texttt{implements} vs.\ \texttt{injects}) and supports deterministic upstream discovery via predecessors and interface-consumer expansion. The LLM-KB schema is \emph{coarser} (single \texttt{depends\_on} relation), which can be sufficient for many dependency summaries but makes higher-precision traversal (e.g., inheritance-only expansion or interface boundary fixes) harder without additional post-processing.

\begin{table*}[t]
\caption{Graph Schema Summary: Node and Edge Types (DKB vs.\ LLM-KB)}
\label{tab:graph_schema}
\centering
\begingroup
\setlength{\tabcolsep}{3pt} % tighten column padding
\renewcommand{\arraystretch}{1.08} % a bit more row breathing room
\scriptsize % slightly smaller to avoid overflow
\begin{tabular}{p{0.12\textwidth} p{0.14\textwidth} p{0.16\textwidth} p{0.20\textwidth} p{0.30\textwidth}}
\toprule
\textbf{System} & \textbf{Element} & \textbf{Type / Label} & \textbf{Direction} & \textbf{Extraction / Semantics} \\
\midrule
DKB & Node & \texttt{class} & -- & Added when a \texttt{class\_declaration} is discovered; node stores \texttt{path} to defining file. \\
DKB & Node & \texttt{interface} & -- & Added when an \texttt{interface\_declaration} is discovered; node stores \texttt{path}. \\
DKB & Node & \texttt{enum} & -- & Added when an \texttt{enum\_declaration} is discovered; node stores \texttt{path}. \\
DKB & Node & \texttt{record} & -- & Added when a \texttt{record\_declaration} is discovered; node stores \texttt{path}. \\
DKB & Node & \texttt{annotation} & -- & Added when an \texttt{annotation\_type\_declaration} is discovered; node stores \texttt{path}. \\
\midrule
DKB & Edge & \texttt{injects} & $A \rightarrow B$ & Emitted when $A$ has a field type or constructor parameter type matching project-local type $B$ (lightweight DI/composition signal). \\
DKB & Edge & \texttt{extends} & $A \rightarrow B$ & Emitted when $A$ declares superclass $B$ (class inheritance) or when an interface extends another interface. \\
DKB & Edge & \texttt{implements} & $A \rightarrow I$ & Emitted when class/record/enum $A$ declares interface $I$ in its implements list. \\
\midrule
DKB & Expansion rule & \texttt{Succ} & -- & Downstream neighborhood: include successors of retrieved seed types up to hop depth $d$. \\
DKB & Expansion rule & \texttt{Pred} & -- & Upstream neighborhood: include predecessors (consumers) of retrieved seed types up to hop depth $d$. \\
DKB & Expansion rule & \texttt{InterfaceConsumerExpand} & -- & If seed type $A$ implements interface $I$, add predecessors of $I$ (interface consumers) to recover upstream controllers/services across interface boundaries. \\
\midrule
LLM-KB & Node & \texttt{class} & -- & Added from LLM-emitted \texttt{class\_name} (per-file structured analysis record). \\
LLM-KB & Edge & \texttt{depends\_on} & $A \rightarrow B$ & Emitted for each dependency string/class $B$ returned by the LLM for class $A$; schema does not distinguish inheritance vs.\ usage unless post-processed. \\
\midrule
LLM-KB & Completeness signal & \texttt{SKIPPED/MISSED} & -- & Logged when an input file in a prompt batch does not appear in the model's structured output (probabilistic extraction incompleteness). \\
\bottomrule
\end{tabular}
\endgroup
\end{table*}

\end{document}